\documentclass[journal=mamobx,manuscript=article,layout=traditional]{achemso}
\setkeys{acs}{articletitle = true}

\usepackage{graphicx}
\graphicspath{{IMAGES/}}

\usepackage{dcolumn}
\usepackage{bm}
\usepackage[utf8x]{inputenc}
\usepackage{lipsum}
\usepackage{resizegather}
\usepackage{amssymb}
\usepackage{amsmath}
\usepackage{booktabs}
\usepackage{xcolor}
\usepackage[symbol]{footmisc}
\usepackage[normalem]{ulem} 

\author{Valerio Sorichetti}
\affiliation{These authors contributed equally}
\alsoaffiliation{Laboratoire de Physique Théorique et Modèles Statistiques (LPTMS), CNRS, Université Paris-Saclay, F-91405 Orsay, France}
\alsoaffiliation{Laboratoire Charles Coulomb (L2C), Univ. Montpellier, CNRS, F-34095, Montpellier, France}
\alsoaffiliation{IATE, INRAE, CIRAD, Montpellier SupAgro, Univ. Montpellier,  F-34060, Montpellier, France}
\email{valerio.sorichetti@universite-paris-saclay.fr}

\author{Andrea Ninarello}
\affiliation{These authors contributed equally}
\alsoaffiliation{CNR-ISC Uos Sapienza, Piazzale A. Moro 2, IT-00185 Roma, Italy}
\alsoaffiliation{Department of Physics, {\textit Sapienza} Universit\`a di Roma, Piazzale A. Moro 2, IT-00185 Roma, Italy}

\author{Jos\'e M. Ruiz-Franco}
\affiliation{CNR-ISC Uos Sapienza, Piazzale A. Moro 2, IT-00185 Roma, Italy}
\alsoaffiliation{Department of Physics, {\textit Sapienza} Universit\`a di Roma, Piazzale A. Moro 2, IT-00185 Roma, Italy}

\author{Virginie Hugouvieux}
\affiliation{IATE, INRAE, CIRAD, Montpellier SupAgro, Univ. Montpellier,  F-34060, Montpellier, France}

\author{Walter Kob}
\affiliation{Laboratoire Charles Coulomb (L2C), Univ. Montpellier, CNRS, F-34095, Montpellier, France}
\alsoaffiliation{Institut Universitaire de France}

\author{Emanuela Zaccarelli}
\affiliation{CNR-ISC Uos Sapienza, Piazzale A. Moro 2, IT-00185 Roma, Italy}
\alsoaffiliation{Department of Physics, {\textit Sapienza} Universit\`a di Roma, Piazzale A. Moro 2, IT-00185 Roma, Italy}

\author{Lorenzo Rovigatti}
\affiliation{Department of Physics, {\textit Sapienza} Universit\`a di Roma, Piazzale A. Moro 2, IT-00185 Roma, Italy}
\alsoaffiliation{CNR-ISC Uos Sapienza, Piazzale A. Moro 2, IT-00185 Roma, Italy}

\title{The effect of chain polydispersity on the elasticity of disordered polymer networks}

\SectionNumbersOn

\begin{document}


\begin{abstract}
Due to their unique structural and mechanical properties, randomly-crosslinked polymer networks play an important role in many different fields, ranging from cellular biology to industrial processes. In order to elucidate how these properties are controlled by the physical details of the  network (\textit{e.g.} chain-length and end-to-end distributions), we generate disordered phantom networks with different crosslinker concentrations $C$ and initial density $\rho_{\rm init}$ and evaluate their elastic properties. We find that the shear modulus computed at the same strand concentration for networks with the same $C$, which determines the number of chains and the chain-length distribution, depends strongly on the preparation protocol of the network, here controlled by $\rho_{\rm init}$. We rationalise this dependence by employing a generic stress-strain relation for polymer networks that does not rely on the specific form of the polymer end-to-end distance distribution. We find that the shear modulus of the networks is a non-monotonic function of the density of elastically-active strands, and that this behaviour has a purely entropic origin.
Our results show that if short chains are abundant, as it is always the case for randomly-crosslinked polymer networks, the knowledge of the exact chain conformation distribution is essential for predicting correctly the elastic properties. Finally, we apply our theoretical approach to published experimental data, qualitatively confirming our interpretations.
\end{abstract}

\maketitle

\section{Introduction}\label{sec:introduction}

For many applications, the elasticity of a crosslinked polymer network is one of its most important macroscopic properties \cite{treloar1975physics}. It is thus not surprising that a lot of effort has been devoted to understand how the features of the network, such as the fraction and functionality of crosslinkers or the details of the microscopic interactions between chain segments, contribute to generate its elastic response~\cite{treloar1943elasticity1,treloar1943elasticity2,flory1943statistical1,flory1943statistical2,flory1985molecular,broedersz2014modeling}. The macroscopic behaviour of a real polymer network (be it a rubber or a hydrogel) depend on many quantities, such as the properties of the polymer and of the solvent, the synthesis protocol and the thermodynamic parameters. However, in experiments it is difficult to disentangle how these different elements contribute to the elastic properties of the material. This task becomes easier in simulations, since all the relevant parameters can be controlled in detail~\cite{grest1990statistical,duering1991relaxation,grest1992kinetics,duering1994structure,everaers1996topological,kenkare1998discontinuous,everaers1999entanglement,kenkare2000theory,svaneborg2008connectivity,gula2020computational}. In this regard, an important feature of real polymer networks that can be exploited is that their elasticity can be described approximately as the sum of two contributions: one due to the crosslinkers and one due to the entanglements \cite{rubinstein1997nonaffine,rubinstein2002elasticity,svaneborg2008connectivity,gula2020computational}. The former can be approximated well by the elastic contribution of the corresponding phantom network \cite{mark2007physical}, i.e. when the excluded volume between the strands is not taken into account.~\cite{svaneborg2008connectivity,gula2020computational} It is therefore very important to understand the role of the chain conformation distribution on the dynamics and elasticity of phantom polymer models.

The distribution of the chemical lengths of the strands between two crosslinkers, \textit{i.e.} the chains, in a network (\textit{chain-length distribution} for short) depends on the chemical details and on the synthesis protocol. For example, in randomly crosslinked networks this distribution is typically exponential \cite{grest1990statistical,higgs1988polydisperse}, whereas chains are monodispersed when end-crosslinking is performed. Regardless of the synthesis route, the presence of short or stretched chains is common, although the exact form of the chain conformation fluctuations is highly non-trivial.
From a theoretical viewpoint, however, the majority of the results on the elasticity of polymer networks have been obtained within the mean-field realm, in which scaling assumptions and chain Gaussianity are assumed~\cite{rubinstein2003polymer,mark2007physical}.
Therefore, simulations can be extremely helpful to clarify the exact role played by the chain-length distribution and better understand experimental results. However, most simulation studies have focused on melt densities, where random or end-crosslinking can be employed efficiently~\cite{grest1990statistical,duering1991relaxation,grest1992kinetics,duering1994structure,kenkare1998discontinuous,kenkare2000theory,svaneborg2008connectivity,gula2020computational}, or have employed idealised lattice networks \cite{everaers1995test,everaers1996topological,escobedo1997simulation,everaers1999entanglement,escobedo1999molecular}. This makes it challenging to compare the results from such simulations with common experimental systems such as hydrogels, which are both low-density and disordered~\cite{richbourg2020swollen}. 

In the present paper, we show that the knowledge of the exact chain end-to-end distribution is essential to correctly predict the linear elastic response of low-density polymer networks. We do so by simulating disordered phantom networks generated with different crosslinker concentration $C$ and initial monomer density $\rho_{\rm init}$. In our systems, the former parameter controls the number of chains and the chain-length distribution, while the latter determines the initial end-to-end distance distribution of the chains and therefore plays a similar role as the solvent quality in an experimental synthesis. To generate the gels we exploit a recently introduced technique based on the self-assembly of patchy particles, which has been proven to correctly reproduce structural properties of experimental microgels~\cite{gnan2017silico, ninarello2019, rovigatti2019numerical}. This method allows us to obtain systems at densities comparable with those of experimental hydrogels, i.e. giving access to swelling regimes inaccessible through the previously employed techniques based on numerical vulcanization of high-density polymer melts~\cite{duering1991relaxation,grest1992kinetics,duering1994structure,everaers1996topological,kenkare1998discontinuous,everaers1999entanglement,svaneborg2008connectivity,gula2020computational}.
We first demonstrate that systems generated with the same $C$ but at different values of $\rho_{\rm init}$ can display very different elastic properties even when probed at the same strand concentration and despite having the same chain length distribution. Secondly, we compare the numerical results to the phantom network theory \cite{mark2007physical}. In order to do so, we determine the theoretical relation between the shear modulus $G$ and the single-chain entropy for generic non-Gaussian chains.
We find a good agreement between theory and simulation only for the case in which the exact chain end-to-end distribution is given as an input to the
theory, with some quantitative deviations appearing at low densities. On the other hand, assuming a Gaussian behaviour of the chains leads to qualitatively wrong predictions for all the investigated systems except the highest-density ones.
Overall, our analysis shows that for low-density polymer networks and in the presence of short chains the knowledge of the exact chain conformational fluctuations is crucial to predict the system elastic properties reliably. Notably, we validate our approach against recently published experimental data~\cite{hoshino2018network,matsuda2019fabrication}, showing that the behaviour of systems where short chains are present cannot be modelled without a precise knowledge of the chain-size-dependent end-to-end distribution.

\section{Theoretical background}\label{sec:theory}

In this section we review some theoretical results on the elasticity of polymer networks, for the most part available in the literature~\cite{treloar1975physics,rubinstein2003polymer,mark2007physical}, by re-organizing them and introducing the terminology and notation that will be employed in the rest of the paper.
We will consider a polydisperse polymer network made of crosslinkers of valence $\phi$ connected by $N_s$ strands. Here and in the following we will assume the network to be composed of $N_s$ elastically-active strands, defined as strands with the two ends connected to distinct crosslinkers, \textit{i.e.}, that are neither dangling ends nor closed loops (\textit{e.g.} loops of order one). Moreover, for those strands which are part of higher-order loops, we assume their elasticity to be independent of the loop order (see \citet{zhong2016quantifying} and \citet{lin2019revisiting}).
We will focus on evaluating the shear modulus $G$ of the gel, which relates a pure-shear strain to the corresponding stress in the linear elastic regime~\cite{landau1970elasticity}. One can theoretically compute $G$ by considering uniaxial deformations of strain $\lambda$ along, for instance, the $x$ axis. We assume the system to be isotropic; moreover, since we are interested in systems with no excluded volume interactions, we assume a volume-preserving transformation \footnote{In the absence of excluded volume interactions the pressure of the system is negative and would therefore collapse if the volume was not kept constant.}, \textit{i.e.}, $\lambda_x = \lambda$ and $\lambda_y = \lambda_z = \lambda^{-1/2}$ as extents of deformation along the three axes.

The starting point to calculate the shear modulus is the single chain entropy, which is a function on the chain's end-to-end distance~\cite{rubinstein2003polymer}. In general, we can write the instantaneous end-to-end vector of a single chain which connects any two crosslinkers as $ \mathbf r(t) = \mathbf R + \mathbf u(t)$, where $\mathbf R \equiv \overline{\mathbf r(t)}$ represents the time-averaged end-to-end vector and $\mathbf u(t)$ the fluctuation term. We also assume that there are no excluded volume interactions, so that the chains can freely cross each other. We thus have $\overline{r^2} = R^2 + \overline{u^2}$, \footnote[2]{Here and in the following, italic is used to indicate the magnitude of the vector.} since $\overline{\mathbf R \cdot \mathbf u(t)} = 0$,  the position and fluctuations of crosslinkers being uncorrelated \cite{mark2007physical}.

The entropy of a chain with end-to-end vector $\mathbf r=(r_x,r_y,r_z)$ is $S_{n}(\mathbf r)  = k_B \log W_{n}(\mathbf r) + A_n$~\cite{flory1976statistical}, where $W_n(\mathbf r)$ is the end-to-end probability density of $\mathbf r$ and $A_n$ is a temperature-dependent parameter that can be set to zero in this context. If the three spatial directions are independent (which is the case, \textit{e.g.}, if $W_{n}(\mathbf r)$ is Gaussian) then $W_{n}(\mathbf r)$ can be written as the product of three functions of $r_x,r_y$, and $r_z$, so that $S_{n}(\mathbf r) = s_n(r_x)+s_n(r_y)+s_n(r_z)$, where $s_n$ is the entropy of a one-dimensional chain. Building upon this result, we can assume that each chain in the network can be replaced by three independent one-dimensional chains parallel to the axes using the so-called \textit{three-chain} approximation~\cite{smith1974modulus,treloar1975physics}. This assumption is exact for Gaussian chains, although for non-Gaussian chains the associated error is small if the strain is not too large~\cite{treloar1975physics}.

We will also assume (i) that the length of each chain in the unstrained state ($\lambda=1$) is ${\tilde r \equiv (\overline{r^2})^{1/2} }\equiv {(R^2+ \overline{u^2})^{1/2}}$, and (ii) that, upon deformation, the chains deform affinely with the network, so that the length of the chain oriented along the $x$ axis becomes $\tilde r_\lambda$ and those of the chains oriented along the $y$ and $z$ axes become $\tilde r_{\lambda^{-1/2}}$. With those assumptions, the single-chain entropy $S_n(\lambda)$ becomes~\cite{treloar1975physics} 

\begin{equation}
S_{n}(\lambda) 
=  \frac{s_n( \tilde r_{\lambda}) + 2 s_{n}( \tilde r_{\lambda^{-1/2}})} 3, 
\label{eq:s_large}
\end{equation}

\noindent
where we need to divide by $3$ since we are replacing each unstrained chain with end-to-end distance $\tilde r$ by three fictitious chains of the same size. Usually, the $\lambda-$dependence of $\tilde r_\lambda$ is controlled by the microscopic model and by the macroscopic conditions (density, temperature, \textit{etc.}). Two well-known limiting cases are the \textit{affine network} model \cite{mark2007physical}, in which both the average positions and fluctuations of the crosslinkers deform affinely, $\tilde r_{\lambda}= \lambda \tilde r$, and the \textit{phantom network} model \cite{mark2007physical}, in which the fluctuations are independent of the extent of the deformation, so that

\begin{equation}
 \tilde r_{\lambda} = {[(\lambda^2 R^2+ \overline{u^2})]^{1/2}}, \ \text{and thus} \ \ \tilde r_{\lambda^{-1/2}} = {[(R^2/\lambda+ \overline{u^2})]^{1/2}}.
 \label{eq:rtilde}
\end{equation}

\noindent The free-energy difference between the deformed and undeformed state of a generic chain is $\Delta F = - T [S_{n}(\lambda) - S_{n}(1)]$ and thus the $x$ component of the tensile force is given by 

\begin{equation}
f_x(\lambda) = \frac{1}{L_{x0}} \frac{d \Delta F}{d \lambda} = -\frac{T}{L_{x0}}  \frac{d S_n(\lambda)}{d \lambda}. 
\label{eq:force}
\end{equation}

\noindent
The latter quantity divided by the section $L_{y0}L_{z0}$ yields the $xx$ component of the stress tensor, which thus reads

\begin{equation}
\sigma_{xx} = -\frac{T R^2}{3 V} \left[ \frac{\lambda}{ \tilde r_{\lambda}} \frac{ds_{n}( \tilde r_{\lambda})}{d  \tilde r_{\lambda}} - \frac{1}{\lambda^2  \tilde r_{\lambda^{-1/2}}} \frac{ds_{n}( \tilde r_{\lambda^{-1/2}})}{d  \tilde r_{\lambda^{-1/2}}} \right],
\end{equation}

\noindent
where we have used Eq.~\eqref{eq:rtilde}.

Since the volume is kept constant, the Poisson ratio is $1/2$ \cite{landau1970elasticity} and hence the single-chain shear modulus $g$ is connected to the Young modulus $Y = \frac{d \sigma_{xx}}{d \lambda}\Bigr|_{\lambda=1}$ by $g = Y / 3$, which implies that

\begin{equation}
\label{eq:g_ph}
g = -\frac{T R^2}{3 V} \left[ \frac{ds_{n}(\tilde r)}{d\tilde r} \left( \frac{1}{\tilde r} - \frac{R^2}{2  \tilde r^3} \right) + \frac{d^2 s_{n}(\tilde r)}{d \tilde r^2} \frac{R^2}{2  \tilde r^2} \right].
\end{equation}

\noindent
We note that, although similar equations can be found in \citet{smith1974modulus} and \citet{treloar1975physics}, to the best of our knowledge Eq.~\eqref{eq:g_ph} has not been reported in the literature in this form. In order to obtain the total shear modulus $G$ of the network, and under the assumption that the effect of higher-order loops can be neglected~\cite{zhong2016quantifying,lin2019revisiting}, one has to sum over the $N_s$ elastically-active chains. Of course, the result will depend on the specific form chosen for the entropy $s_n$. We stress that a closed-form expression of the end-to-end probability density $W_{n}(\mathbf r)$ is not needed, since only its derivatives play a role in the calculation. Hence, it is sufficient to know the force-extension relation for the chain, since, as discussed above, the component of the force along the pulling direction satisfies Eq.~\eqref{eq:force} (see also Sec.~\ref{sec:models}).

For a freely-jointed chain (FJC) \cite{rubinstein2003polymer} of $n$ bonds of length $b$, $W_n(\mathbf r)$ has the following form~\cite{jernigan1969distribution,treloar1975physics}:

\begin{equation}	
\label{eq:W_FJC}
W_n(\mathbf r) = \left[ \frac{n(n-1)}{8 \pi r b^2} \right] \sum_{t=0}^\tau \frac{(-1)^t}{t!(n - t)!} \left[ \left( \frac{nb - r}{2b} \right) - t \right]^{n-2},
\end{equation}

\noindent
where $\tau = \lfloor (nb - r) / 2b \rfloor$, \textit{i.e.}, the largest integer smaller than $(nb - r) / 2b$. 

In the limit of large $n$, Eq.~\eqref{eq:W_FJC} reduces to a Gaussian \cite{jernigan1969distribution}:

\begin{equation}
\label{eq:W_G}
W_n^G(\mathbf r) = \left( \frac{3}{2 \pi n b^2} \right)^{3/2} \exp\left( - \frac{3 r^2}{2 n b^2} \right).
\end{equation}

\noindent
Under this approximation, the shear modulus takes the well-known form

\begin{equation}
\label{eq:G_gaussian}
G^{G} = \frac{k_B T}{V} \sum_{i}^{N_s}  \frac{R_i^2}{n_ib^2} = \left\langle \frac{R^2}{nb^2} \right\rangle k_B T \nu \equiv A k_B T \nu,
\end{equation}

\noindent
where $\nu = N_s / V$ is the number density of elastically-active strands and $A$ is often called the \textit{front factor}~\cite{james1943theory,tobolsky1961rubber,smith1974modulus,toda2018rubber}. We have also introduced the notation $\langle \cdot \rangle = N_s^{-1} \sum_i^{N_s}\cdot$ for the average over all the strands in the system. In the particular case that the $\overline{r_i^2}$ values of the different chains are Gaussian distributed (a distinct assumption from the one that $W_n(\mathbf r)$ is Gaussian), which is the case, for example, for end-crosslinking starting from a melt of precursor chains\cite{duering1994structure}, it can be shown that $A = 1-\frac{2}{\phi}$ (we recall that $\phi$ is the crosslinker valence), so that one obtains the commonly reported expression (see also  Sec.~\ref{sec:shear_gauss}) \cite{rubinstein2003polymer,mark2007physical}

\begin{equation}
\label{eq:G_textbook}
G^G = \left(1 - \frac 2 \phi \right) k_B T \nu.
\end{equation}

\noindent
Eq.~\eqref{eq:G_gaussian} was derived from Eq.~\eqref{eq:g_ph}, which assumes the validity of the phantom network model. If one assumes, on the other hand, that the affine network model is valid, a different expression for $G$ is obtained (see Supplementary material).

To obtain a more accurate description of the end-to-end probability distribution for strained polymer networks, one has to go beyond the Gaussian model and introduce more refined theoretical assumptions. Amongst other approaches, the Langevin-FJC \cite{treloar1975physics} (L-FJC), the extensible-FJC \cite{fiasconaro2019analytical} (ex-FJC) and the worm-like chain \cite{petrosyan2017improved} (WLC) have been extensively used in the literature. In the L-FJC model the force-extension relation is approximated using an inverse Langevin function, whereas in the ex-FJC model bonds are modeled as harmonic springs. These models give a better description of the system's elasticity when large deformations are considered. The WLC model, in which chains are represented as continuously-flexible rods, is useful when modeling polymers with high persistence length (compared to the Kuhn length). More details about these models can be found in Sec.~\ref{sec:models}.

\section{Model and Methods}\label{sec:model_methods}

We build the polymer networks by employing the method reported in \citet{gnan2017silico}, which makes use of the self-assembly of a binary mixture of limited-valence particles. Particles of species $A$ can form up to four bonds (valence $\phi=4$) and bond only to $B$ particles only, thus acting as crosslinkers. Particles of species $B$ can form up to two bonds ($\phi=2$) and can bond to $A$ and $B$ particles. We carry out the assembly of $N_{\rm init} = N_A + N_B = 5 \cdot 10^4$ particles at different number density $\rho_{\rm init} = N_{\rm init}/V$, with $V$ the volume of the simulation box, and different crosslinker concentration $C = N_A / (N_B + N_A)$. We consider two initial densities $\rho_{\rm init} = 0.1, 0.85$,  and $C = 1\%$, $5\%$ and $10\%$. The results are averaged over two system realizations for each pair of $\rho_{\rm init}, C$ values.

The assembly proceeds until an almost fully-bonded percolating network is attained, \textit{i.e.}  the fraction of formed bonds is at least $N_{\rm bond}/N_{\rm bond}^{\rm max}=99.9\%$, where $N_{\rm bond}^{\rm max}={(4N_A+2N_B)/2}$ is the maximum number of bonds. The self-assembly process is greatly accelerated thanks to an efficient bond-swapping mechanism~\cite{sciortino2017three}. When the desired fraction $N_{\rm bond}/N_{\rm bond}^{\rm max}$ is reached, we stop the assembly, identify the percolating network and remove all particles or clusters that do not belong to it. Since some particles are removed, at the end of the procedure the values of $\rho_{\rm init}$ and $C$ change slightly. However, these changes are small (at most $10\%$) and in the following we will hence use the nominal (initial) values of $\rho_{\rm init}$ and $C$ to refer to the different networks.

\begin{figure}
\includegraphics[width=0.5\textwidth]{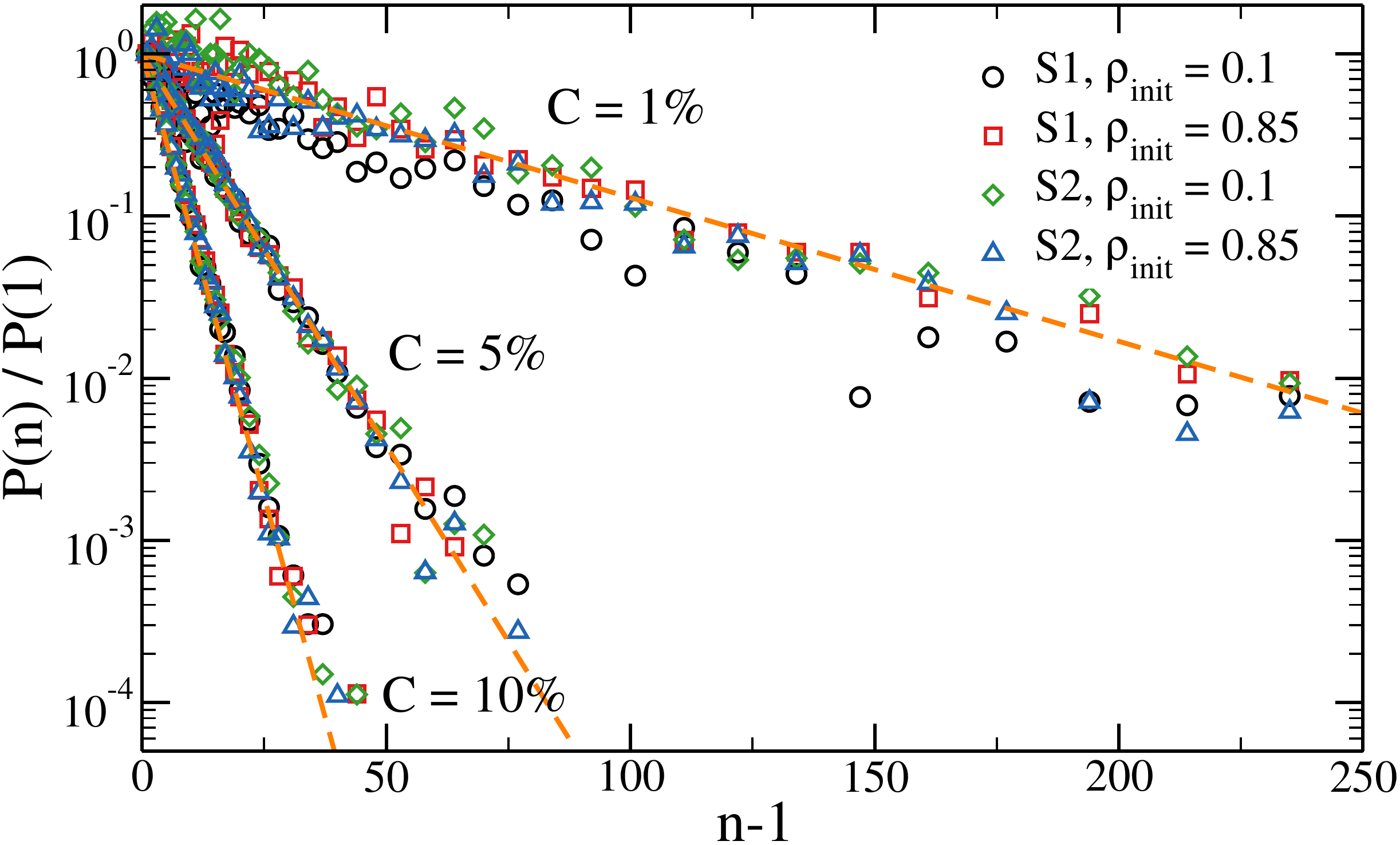}
\caption{\label{fig:histo_N} Rescaled distribution of chain lengths for all the simulated systems. We report the data for two samples (S1 and S2) generated with two values of the initial density each. The orange dashed lines are the theoretical prediction of Eq.~\eqref{eq:chailength}.}
\end{figure}

The normalised distribution of the chemical lengths $n$ of the chains, $P(n)/P(1)$, that constitute the network is shown in Figure~\ref{fig:histo_N}. Here the chemical chain length is here defined as the number of particles in a chain, excluding the crosslinkers, so that a chain with ${n+1}$ bonds has length $n$ \footnote{We recall that, since two crosslinkers cannot bind to each other, the minimum chain length is $n=1$ (for $2$ bonds).}. In all cases the distribution decays exponentially, as it is also the case for random-crosslinking from a melt of precursor chains~\cite{grest1990statistical}. Moreover, $P(n)$ does not depend on the initial density~\cite{gnan2017silico,rovigatti2017internal} and, as one expects given the equilibrium nature of the assembly protocol, it is fully reproducible. This distribution can be estimated from the nominal values of $\phi$ and $C$ \textit{via} the well-known formula of Flory~\cite{flory1953principles}:

\begin{equation}
\label{eq:chailength}
\frac{P(n)}{P(1)} = \left( 1 - \frac{1}{\langle n \rangle} \right)^{n - 1},
\end{equation}

\noindent
where $\langle n \rangle  = 2(1-C)/\phi C$ is the mean chain length~\cite{rovigatti2017internal} which, using the nominal crosslinker valence ($\phi = 4$) and concentrations, takes the values $49.5$, $9.5$ and $4.5$ for $C = 1\%$, $5\%$ and $10\%$, respectively. The parameter-free theoretical probability distribution is shown as orange dashed lines in Fig.~\ref{fig:histo_N} and reproduces almost perfectly the numerical data.

The network contains a few defects in the form of dangling ends (chains which are connected to the percolating network by one crosslinker only) and first-order loops, that is, chains having both ends connected to the same crosslinker\cite{lin2019revisiting}. Since there are no excluded volume interactions, these defects are elastically inactive and therefore do not influence the elastic properties of the network~\cite{lang2018elasticity,panyukov2019loops,lin2019revisiting}. For the configurations assembled at $C=1\%$, the percentage of particles belonging to the dangling ends is $\approx 10\%$ for $\rho_{\rm init}=0.1$ and $\approx 6\%$ for $\rho_{\rm init}=0.85$. For higher values of $C$, the percentages are much smaller (\textit{e.g.} $\approx 2\%$ for $C=5\%,\rho_{\rm init}=0.1$ and  $\approx 1\%$ for $C=10\%,\rho_{\rm init}=0.1$).
In order to obtain an ideal fully-bonded network, the dangling ends are removed. We note that during this procedure, the crosslinkers connected to dangling ends have their valence reduced from $\phi=4$ to $\phi=3$ or $2$ (in the latter case, they become type $B$ particles). The percentage of so-created $3$-valent crosslinkers remains small: For $\rho_{\rm init}=0.1$ is $\approx 15\%, 4\%$, and $2\%$ for $C=1\%,5\%$, and $10\%$ respectively. The presence of these crosslinkers slightly changes the average crosslinker valence, but does not influence the main results of this work. 

Once the network is formed, we change the interaction potential, making the bonds permanent and thus fixing the topology of the network. Since we are interested in understanding the roles that topology and chain size distribution of a polymer network play in determining its elasticity, we consider interactions only between bonded neighbours, similarly to what has been done in \citet{duering1994structure}. Particles that do not share a bond do not feel any mutual interaction, and hence chains can freely cross each other (whence the name \textit{phantom network}). Two bonded particles interact through the widely used Kremer-Grest potential \cite{kremer1990dynamics}, which is given by the sum of the Weeks-Chandler-Andersen (WCA) potential \cite{weeks1971role},

\begin{equation}\label{wca}
\mathcal{U}_{\rm WCA}(r)=\begin{cases}
    4\epsilon\left[\left(\frac{\sigma}{r}\right)^{12}-\left(\frac{\sigma}{r}\right)^{6}\right] + \epsilon  & \text{if $r \le 2^{\frac{1}{6}}\sigma$}\\
    0 &  \text{otherwise}
  \end{cases},
\end{equation}

\noindent which models steric interactions, and of a finite extensible nonlinear elastic (FENE) potential, \textit{i.e.}, 

\begin{equation}
\mathcal{U}_{\rm FENE}(r)=- \frac{k r_0^2} 2 \ln\left[1-\left(\frac{r}{r_0}\right)^2\right],
\end{equation}

\noindent
which models the bonds. We set $k= 30 \epsilon/\sigma^2$ and $r_0 = 1.5 \sigma$. Here and in the following, all quantities are given in reduced units. The units of energy, length and mass are respectively $\epsilon$, $\sigma$ and $m$, where $\epsilon$, and $\sigma$ are defined by Eq.~\eqref{wca} and $m$ is the mass of a particle, which is the same for $A$ and $B$ particles. The units of temperature, density, time and elastic moduli are respectively $[T]=\epsilon/k_B,[\rho]= \sigma^{-3}$, $[t]=\sqrt{m \sigma^2/\epsilon}$, and $[G]=\epsilon \sigma^{-3}$. In these units, the Kuhn length of the model is $b = 0.97$ \cite{kremer1990dynamics}.

We run molecular dynamics simulations in the $NVT$ ensemble at constant temperature $T=1.0$ by employing a Nos\'e-Hoover thermostat~\cite{martyna1992nose}. Simulations are carried out using the LAMMPS simulation package~\cite{plimpton1993fast}, with a simulation time step $\delta t=0.003$.

In order to study the effects of the density on the elastic properties, the initial configurations are slowly and isotropically compressed or expanded to reach the target densities $\rho=0.1, 0.2, 0.5, 0.85, 1.5$. Then, a short annealing of $10^6$ steps and subsequently a production run of $10^7$ steps are carried out. Even for the system with the longest chains, the mean-squared displacement of the single particles reaches a plateau, indicating that the chains have equilibrated (see Supplementary material).

For each final density value, we run several simulations for which we perform a uniaxial deformation in the range $\lambda_{\alpha}\in\left[0.8, 1.2\right]$ along a direction $\alpha$, where 
$\lambda_{\alpha}=L_{\alpha}/L_{\alpha,0}$ is the extent of the deformation and $L_{\alpha,0}$ and $L_{\alpha}$ are the initial and final box lengths along $\alpha$, respectively. The deformation is carried out at constant volume with a deformation rate of $10^{-1}$. To confirm that the system is isotropic, we perform the deformation along different spatial directions $\alpha$.

\begin{figure}
\includegraphics[width=0.45\textwidth]{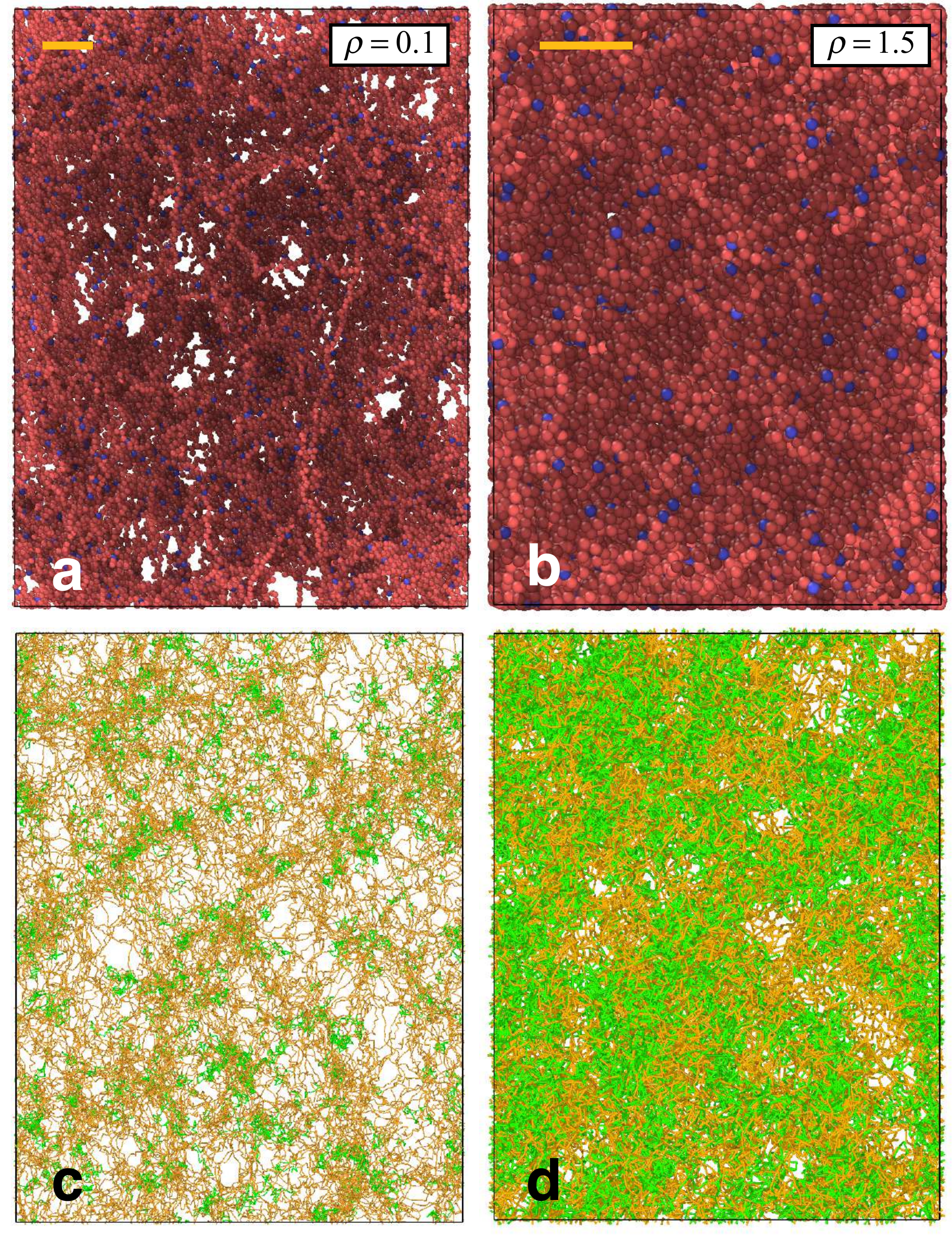}
\caption{Snapshots of a network with crosslinker concentration $C=5\%$ and assembly density $\rho_{\rm init}=0.1$ at (\textbf{a}, \textbf{c}) $\rho = 0.1$ and (\textbf{b}, \textbf{d}) $\rho = 1.5$, subject to a uniaxial deformation along the vertical direction with $\lambda=1.2$ \textbf{a}-\textbf{b}: Turquoise and red particles indicate crosslinkers and monomers, respectively. The orange scale bars in the top left corners are $10 \, \sigma$ long. \textbf{c}-\textbf{d}: The structural response during the uniaxial deformation is represented by the spatial configuration of chains. Orange segments: bonds belonging to overstretched chains ($r>0.95 \cdot nb$); green segments: bonds belonging to unstretched chains ($r<0.95 \cdot nb$).
\label{fig:snapshot}}
\end{figure}

Figure \ref{fig:snapshot} shows representative snapshots of the $C = 5\%$, $\rho_{\rm init}=0.1$ system at low ($\rho=0.1$) and high ($\rho=1.5$) density, subject to a uniaxial deformation along the vertical direction. In panels~\ref{fig:snapshot}\textbf{a}-\textbf{b} we show the particles (monomers and crosslinkers), highlighting the highly disordered nature of the systems and their structural heterogeneity, which is especially evident at low density. The same systems are also shown in panels~\ref{fig:snapshot}\textbf{c}-\textbf{d}, where we use different colours to display bonds of overstretched chains (defined here as chains with $r>0.95 \cdot nb$) and unstretched chains ($r< 0.95 \cdot nb$), highlighting the heterogeneous elastic response of these systems when they are subject to deformations.

Once the system acquires the target value of $\lambda_{z}$, we determine the diagonal elements of the stress tensor $\sigma_{\alpha\alpha}$ and compute the engineering stress $\sigma_{\rm eng}$ as~\cite{rubinstein2003polymer}:

\begin{equation}
\label{eq:true_stress}
\sigma_{\rm eng}=\frac{\sigma_{\rm tr}}{\lambda_{z}}=\frac{1}{\lambda_{z}}\left[\sigma_{zz}-\frac{1}{2}\left(\sigma_{xx}+\sigma_{yy}\right)\right],
\end{equation}

\noindent where $\sigma_{tr}$ is the so-called true stress~\cite{treloar1975physics,doi1996introduction}.The shear modulus $G$ is then the quantity that connects the engineering stress and the strain through the following relation \cite{rubinstein2003polymer}:

\begin{equation}
\label{eq:fitting}
\sigma_{\rm eng}=G\left[\left(\lambda_{z}-\lambda_{\rm ref}\right)-\frac{1}{\left(\lambda_{z}-\lambda_{\rm ref}\right)^{2}}\right]\ .
\end{equation}

\noindent In Eq.~\eqref{eq:fitting} $\lambda_{\rm ref}$ is an extra fit parameter that we add to take into account the fact that in some cases $\sigma_{\rm eng} \ne 0$ for $\lambda_{z}=1$, which signals the presence of some pre-strain in our configurations. The stress-strain curves we use to estimate $G$ are averaged over 10 independent configurations obtained by the randomisation of the particle velocities, prior to deformation, with a Gaussian distribution of mean value $T=1.0\epsilon/k_{B}$ in order to reduce the statistical noise.

\begin{figure}
\includegraphics[width=0.45\textwidth]{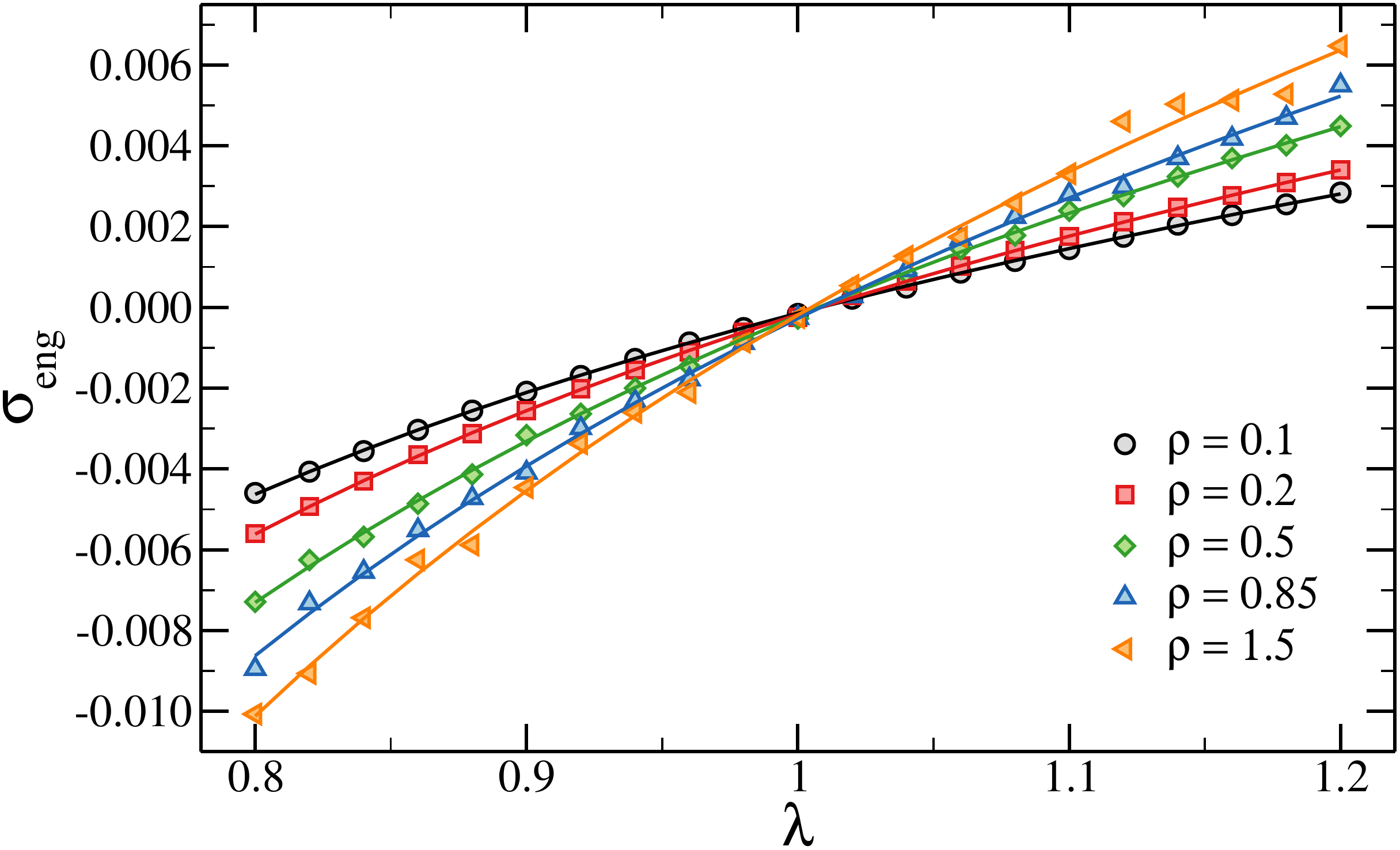}
\caption{\label{fig:stress_strain}Example of stress-strain curves for the $C = 1\%$, $\rho_{\rm init} = 0.1$ system. Symbols are simulation data, lines are fits with Eq.~\eqref{eq:fitting}.}
\end{figure}

Figure~\ref{fig:stress_strain} shows the numerical data for the stress-strain curves for the $C = 1\%$, $\rho_{\rm init} = 0.1$ system. We also report the associated theoretical curves, fitted to Eq.~\eqref{eq:fitting}, through which we obtain an estimate of the shear modulus.

\section{Results and discussion}\label{sec:results}

\begin{figure}
  \includegraphics[width=0.45\textwidth]{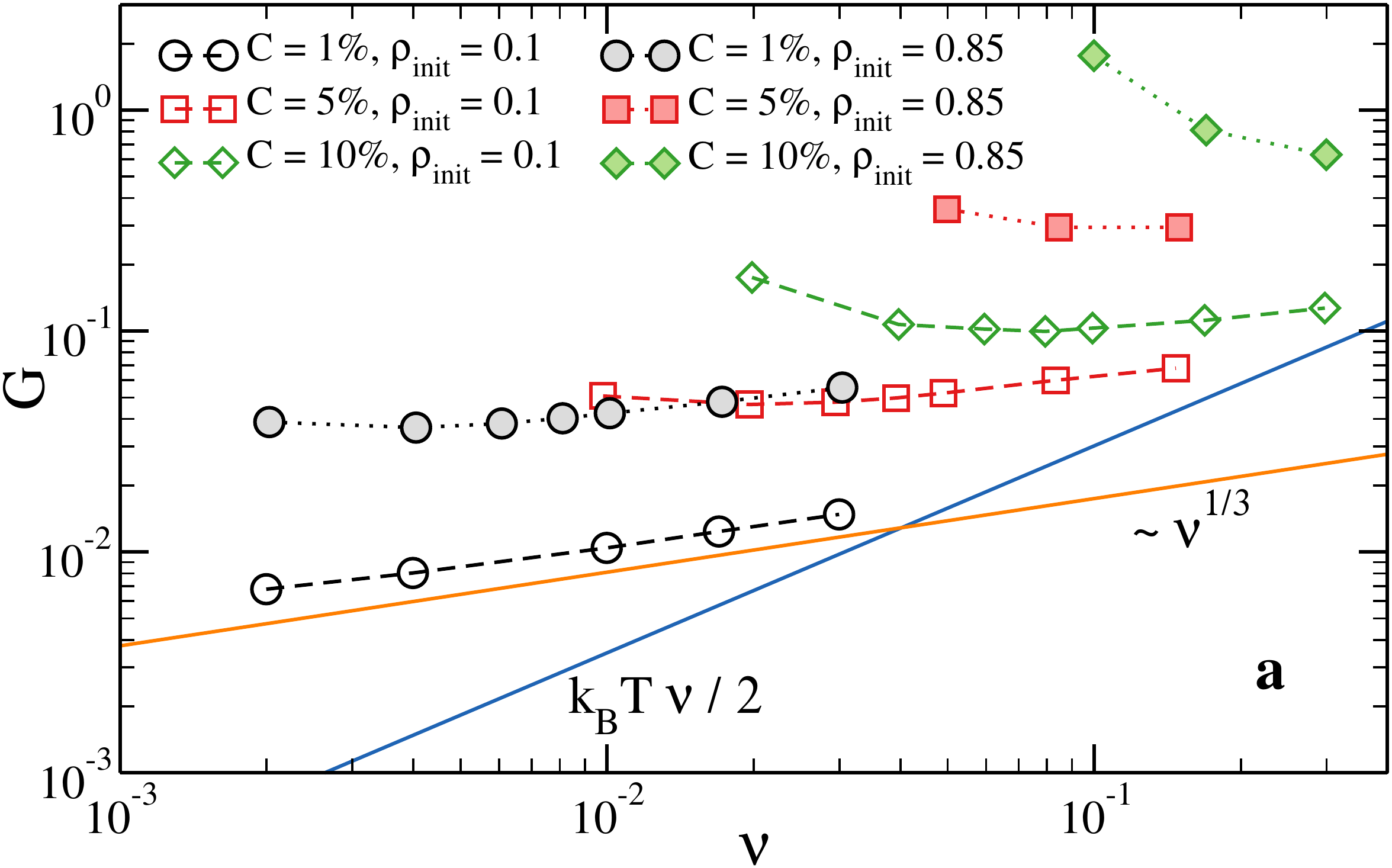}
  \includegraphics[width=0.45\textwidth]{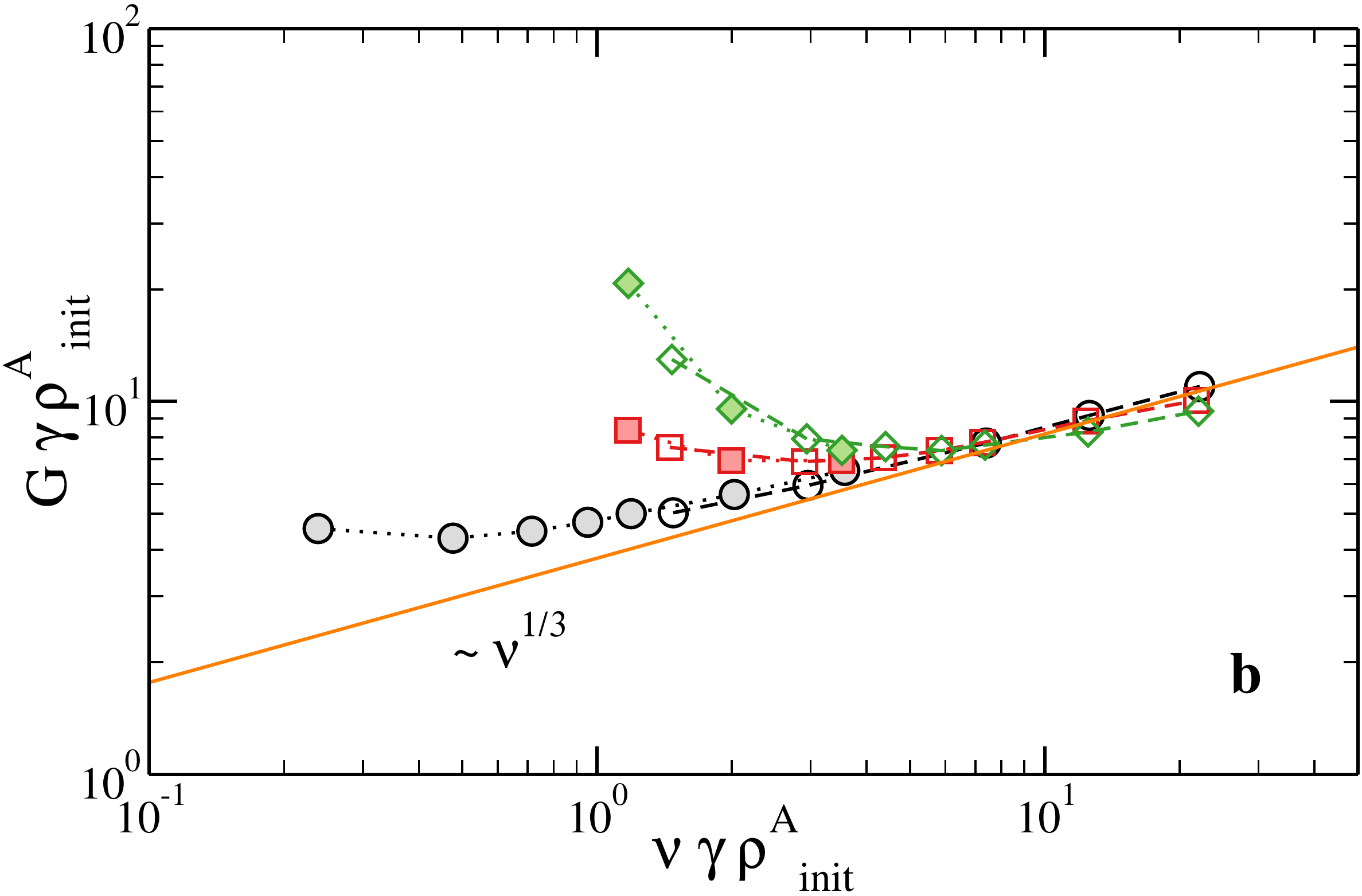}
\caption{\label{fig:G_sim_all}\textbf{a}: Shear modulus as a function of the elastically-active strand number density $\nu$ for all the investigated systems. Solid blue line: Eq.~\eqref{eq:G_textbook} with $\phi=4$. Solid orange line: slope $1/3$. \textbf{b}: Same as \textbf{a}, with both $G$ and $\nu$ rescaled by $\gamma \rho^{A}_{\rm init}$, where $\gamma = 0.74$ for $\rho_{\rm init} = 0.85$ and $\gamma = 1$ for $\rho_{\rm init} = 0.1$ is a fit parameter and $\rho^{A}_{\rm init}= C \rho_{\rm init}$ is the initial crosslinker density.}
\end{figure}

We use the simulation data to estimate $\tilde r \equiv (\overline{r^2})^{1/2}$ (RMS end-to-end distance) and $R \equiv \overline{r}$ for each chain to compute the elastic moduli of the networks through Eq.~\eqref{eq:g_ph}. In the following, we will refer to the elastic moduli computed in this way with the term ``theoretical''.

Figure \ref{fig:G_sim_all}\text{a} shows the shear modulus as computed in simulations for all investigated systems as a function of $\nu$, the density of elastically-active strands. First of all, we observe that systems generated at the same $C$ but with different values of $\rho_{\rm init}$ exhibit markedly different values of the shear modulus when probed under the same conditions (\textit{i.e.} same strand density). This result highlights the fundamental role of crosslinking process, which greatly affects the initial distribution of the chains' end-to-end distances even when the number of chains and their chemical length distribution, being dependent only on $C$ (see Fig.~\ref{fig:histo_N}), are left unaltered. Thus, the echo of the difference between the initial end-to-end distributions gives rise to distinct elastic properties of the phantom networks even when probed at the same strand density. 

In Fig.~\ref{fig:G_sim_all}\text{a} we also plot the behaviour predicted by Eq.~\eqref{eq:G_textbook} (blue line), which assumes Gaussian distributed end-to-end distances. Even though the numerical data seem to approach this limit at very large values of the density, they do so with a slope that is clearly smaller than unity. For the $C = 1\%$ sample this slope is almost exactly $1/3$, while it is very close to this number for the $C = 5\%$ and $C = 10\%$ samples assembled at $\rho_{\rm init} = 0.1$. This behaviour can be understood at the qualitative level from Eq.~\eqref{eq:G_gaussian}: $R$ is the average distance between crosslinkers and therefore it changes affinely upon compression or expansion, thereby scaling as $R \propto \nu^{-1/3}$~\cite{gundogan2002rubber,horkay2000osmotic}. As a result, in the Gaussian limit the shear modulus scales as $G^G \propto \nu^{1/3}$ ~\cite{panyukov1990scaling,horkay2000osmotic,gundogan2002rubber,hoshino2018network}. As discussed above, our results show that the way this limiting regime is approached depends on the crosslinker concentration $C$ and on the preparation state, which is here controlled by $\rho_{\rm init}$.

The quantitative differences in the elastic response of systems with different $C$ and $\rho_{\rm init}$ can be partially rationalised by looking at the scaling properties of the end-to-end distances. We notice that the RMS equilibrium end-to-end distance $R(n)$ of the strands for different values of $\rho_{\rm init}$ and $C$ nearly collapses on a master curve when divided by the initial crosslinker density, $\rho^{A}_{\rm init}= C \rho_{\rm init}$ (see Supplementary material). A slightly better agreement is found if the heuristic factor $\gamma \rho^{A}_{\rm init}$, with $\gamma = 0.74$ for $\rho_{\rm init} = 0.85$ and $\gamma = 1$ for $\rho_{\rm init} = 0.1$ is used. Based on this observation, we rescale the data of Fig.~\ref{fig:G_sim_all}\textbf{a} multiplying both $G$ and $\nu$ by $\gamma \rho^{A}_{\rm init}$. The result is shown in Fig.~\ref{fig:G_sim_all}\textbf{b}: One can see that the shear modulus of systems with the same $C$ but different values of $\rho_{\rm init}$ nicely fall on the same curve. Moreover, in the large-$\nu$ limit, where all the curves tend to have the same slope, a good collapse of the data of systems with different $C$ is also observed. 

The differences arising between systems at different $C$ can be explained by noting that the crosslinker concentration controls the relative abundance of chains with different $n$, whose elastic response cannot be rescaled on top of each other by using $n$ but depends on their specific end-to-end distribution (see \textit{e.g.} Appendix~\ref{sec:models}). As a result, the elasticity of networks generated at different $C$ cannot be rescaled on top of each other. In particular, systems with more crosslinkers, and hence more short chains, will deviate earlier and stronger from the Gaussian behaviour.

\begin{figure}
\includegraphics[width=0.31\textwidth]{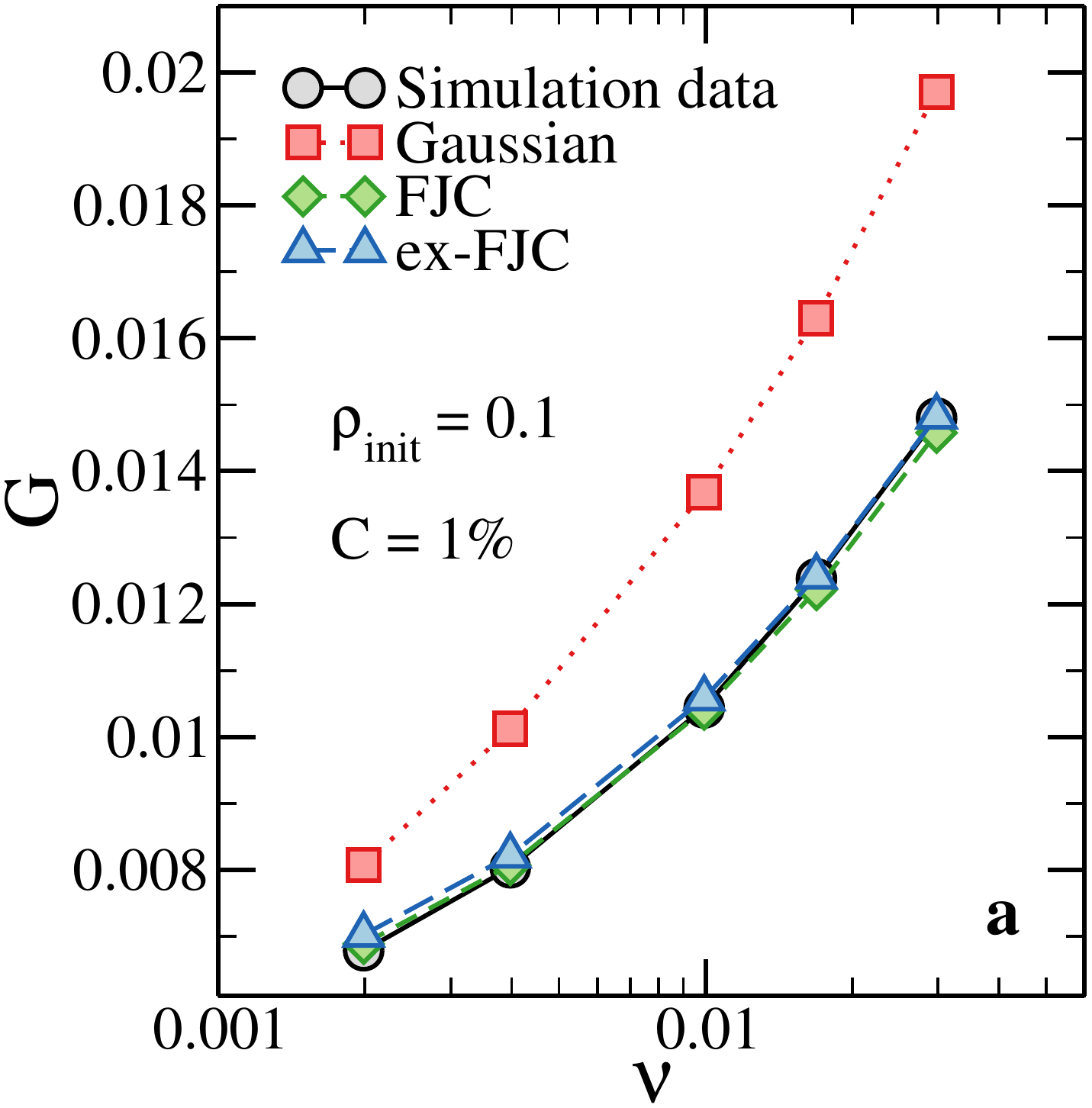}
\includegraphics[width=0.30\textwidth]{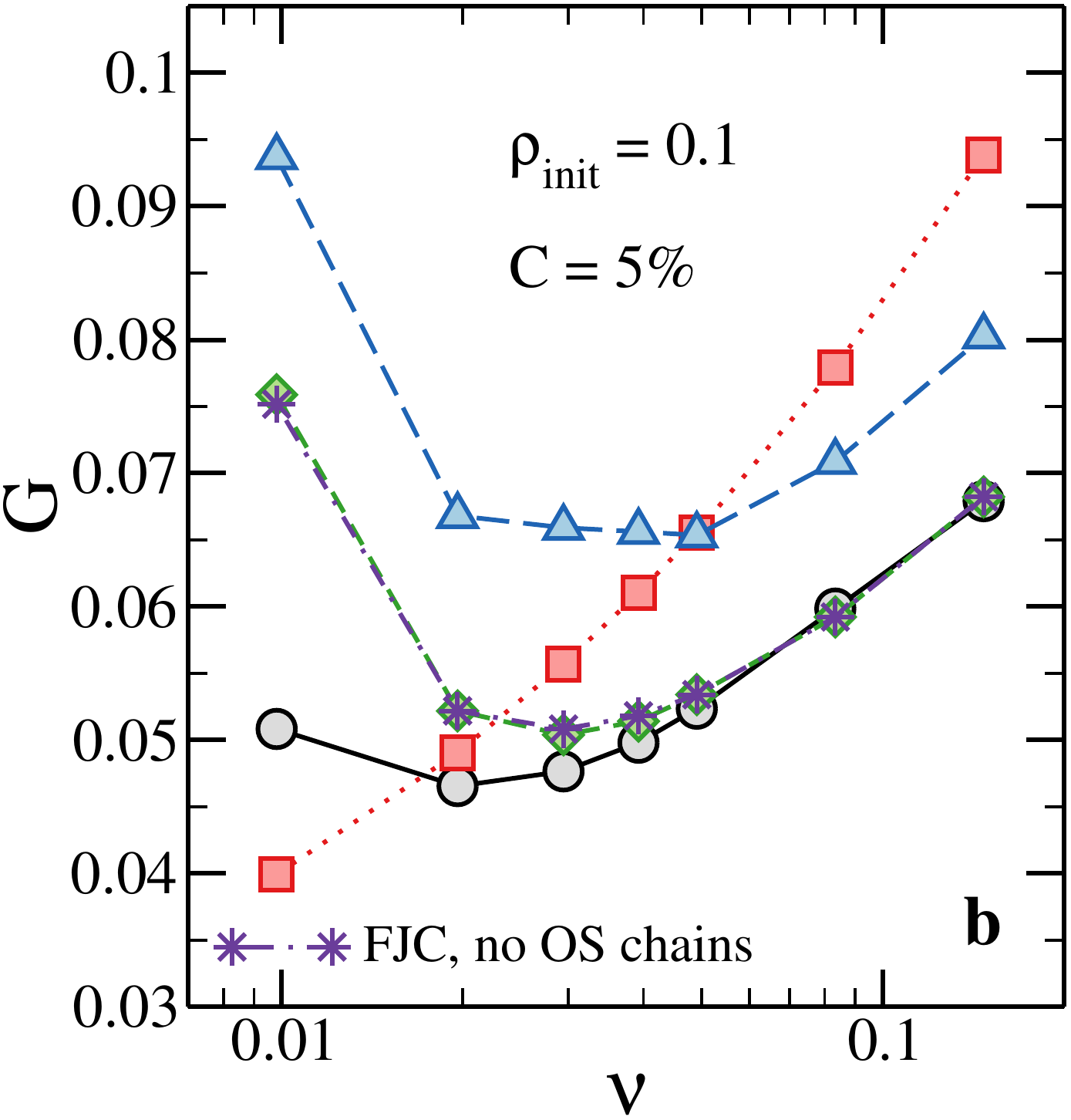}
\includegraphics[width=0.29\textwidth]{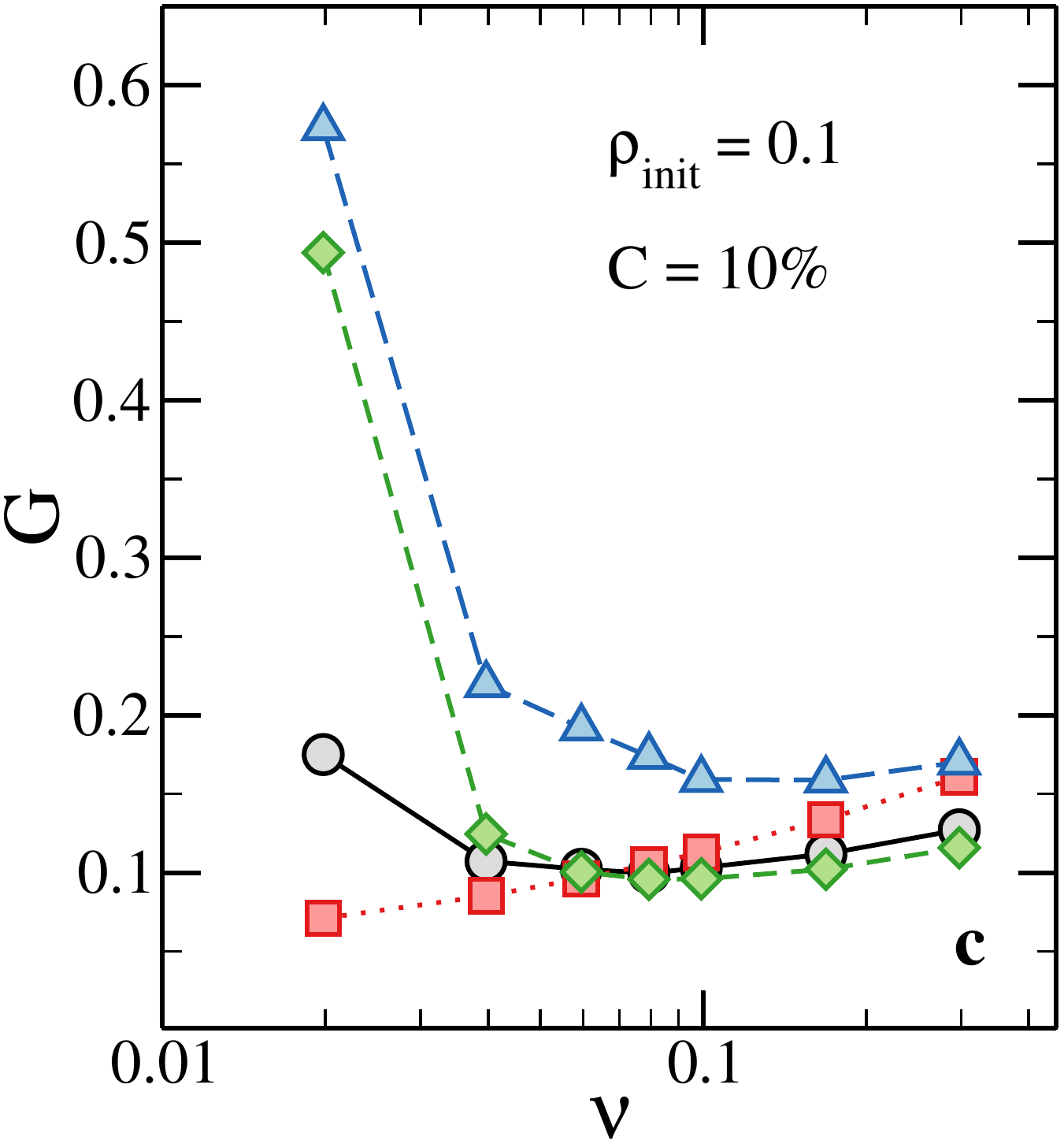}
\includegraphics[width=0.31\textwidth]{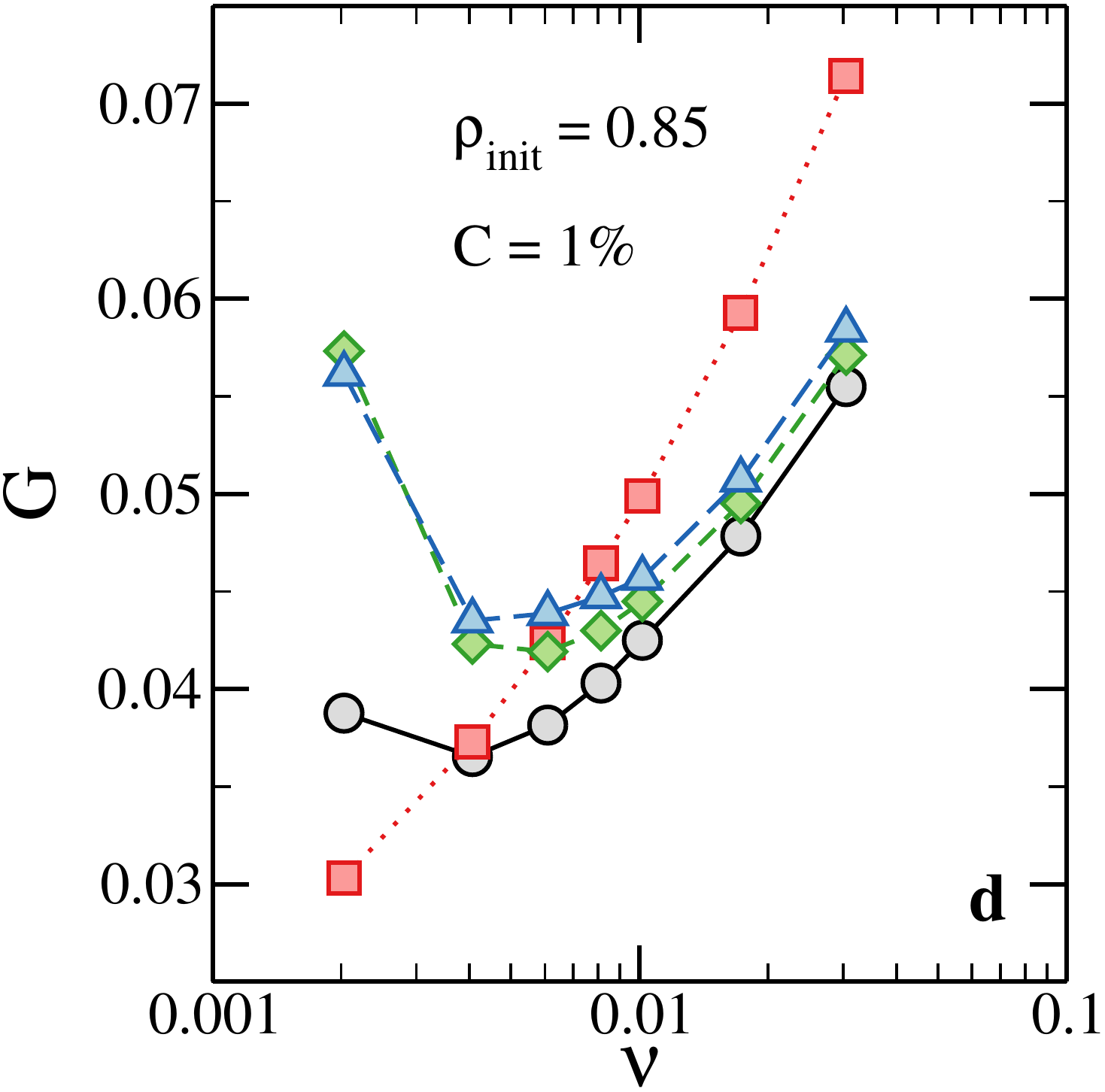}
\includegraphics[width=0.295\textwidth]{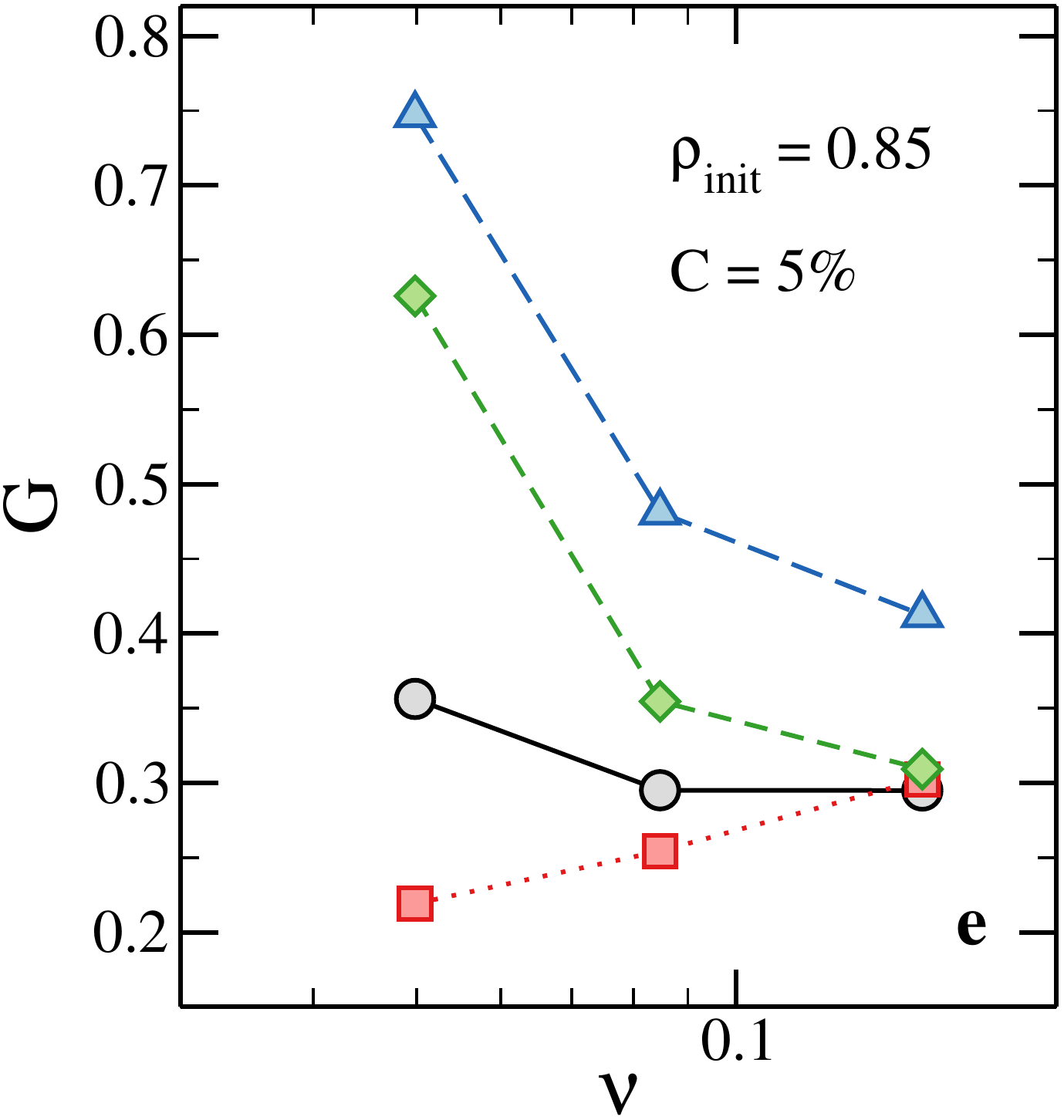}
\includegraphics[width=0.29\textwidth]{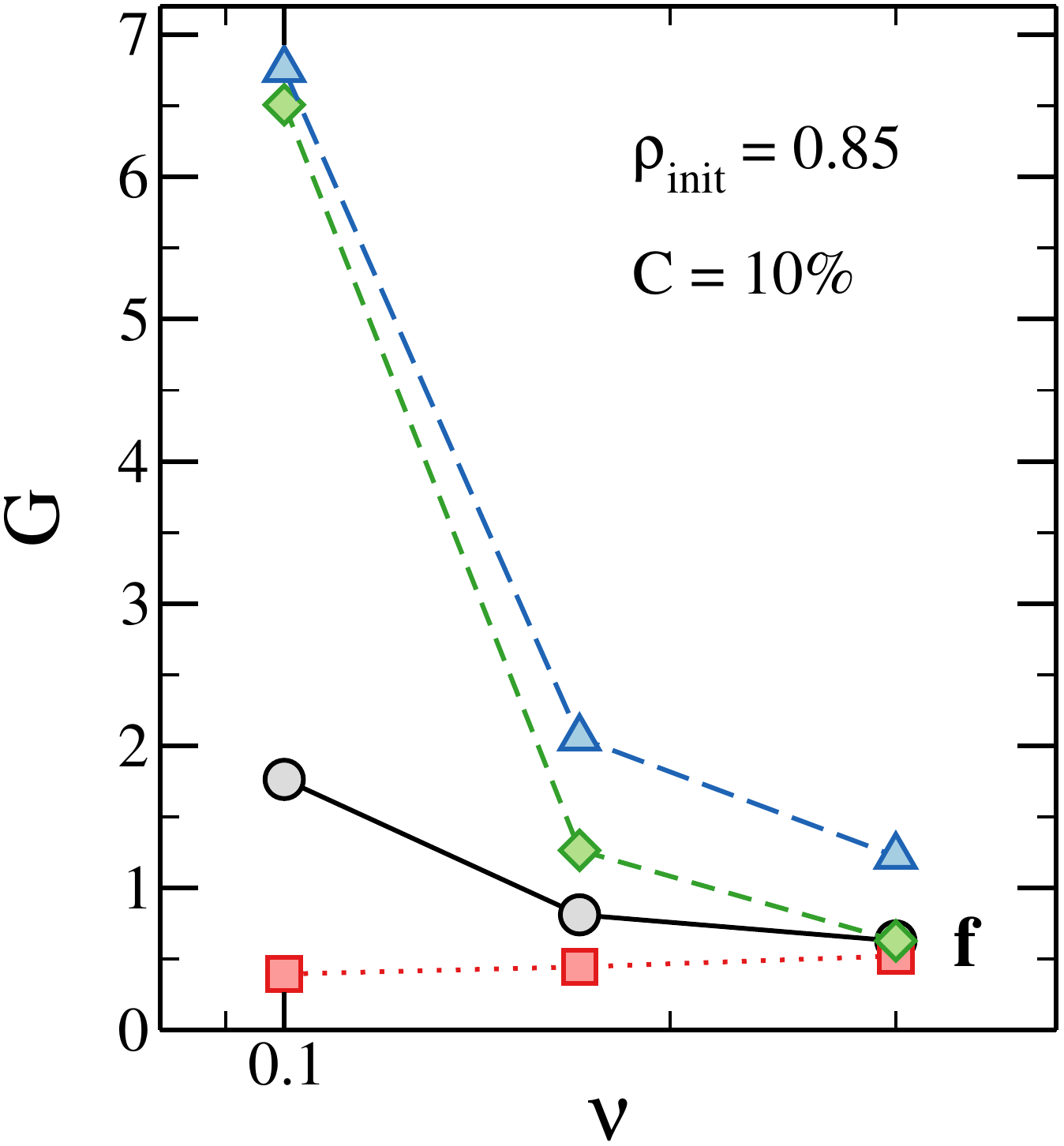}
\caption{\label{fig:G_ratio} Comparison between the shear moduli obtained through Eq.~\eqref{eq:g_ph} and three different approximations and the numerical ones ($G$) for the simulated systems (see legends). Dashed-dotted line / stars in panel b: FJC approximation with no overstretched chains (chains with $r \geq 0.95 \cdot nb$).}
\end{figure}

Interestingly, $G$ exhibits a non-monotonic behaviour as a function of $\nu$; this feature appears for all but the lowest $C$ and $\rho_{\rm init}$ values. This behaviour, which has been also observed in hydrogels~\cite{horkay2000osmotic,gundogan2002rubber,itagaki2010water,hoshino2018network,matsuda2019fabrication}, cannot be explained assuming that the chains are Gaussian, since in this case one has for all $\nu$ that $G\propto \nu^{1/3}$, as discussed above. Given that our model features stretchable bonds, at large strains it cannot be considered a FJC, being more akin to an ex-FJC~\cite{fiasconaro2019analytical}. Therefore, one might be tempted to ascribe the increase of $G$ upon decreasing $\nu$ to the energetic contribution. For this reason, in addition to the Gaussian and FJC descriptions we also plot in Fig.~\ref{fig:G_ratio}b the shear modulus estimated by neglecting the contributions of those chains that have $r \geq 0.95 \cdot nb$ (\textit{i.e.} of the overstretched chains). Since the two sets of data overlap almost perfectly, we confirm that the energetic contribution due to the few overstretched chains is negligible: We can thus conclude that the non-monotonicity we observe has a purely entropic origin. This holds true for all the systems investigated except for the $C = 10\%$, $\rho_{\rm init} = 0.85$ system, which contains the largest number of short, overstretched chains, as can be seen in Figure 1. 

In Fig.~\ref{fig:G_ratio} we compare the numerical shear modulus for all investigated systems with  estimates as predicted by different theories, with the common assumption that the three-chain model remains valid (see Sec.~\ref{sec:theory}). In particular, we show results obtained with the FJC (Eq.~\eqref{eq:W_FJC}), Gaussian (Eq.~\eqref{eq:W_G}), and ex-FJC (see Sec.~\ref{sec:models}) models. One can see that the agreement between the theoretical and numerical results is always better for larger values of $\nu$, i.e. when chains are less stretched. Moreover, the agreement between data and theory is better for systems generated at smaller $\rho_{\rm init}$. We note that the Gaussian approximation, which predicts a monotonically-increasing dependence on $\nu$, fails to reproduce the qualitative behaviour of $G$, whereas the ex-FJC systematically overestimates $G$. The FJC description is the one that consistently achieves the best results, although it fails (dramatically at large $C$) at small densities. We ascribe this qualitative behaviour to the progressive failure of the three-chain assumption as the density decreases. Since the three-chain model is known to overestimate the stress at large strains compared to more complex and realistic approximations such as the tetrahedral model~\cite{treloar1975physics}, the resulting single-chain contribution to the elastic modulus for stretched chains is most likely overestimated as well. Regardless of the specific model used, our results suggest that when the samples are strongly swollen, something that it is possible to achieve in experiments~\cite{hoshino2018network}, any description that attempts to model the network as a set of independent chains gives rise to a unreliable estimate of the overall elasticity even when energetic contributions due to stretched bonds do not play a role.

\begin{figure}
\includegraphics[width=0.6\textwidth]{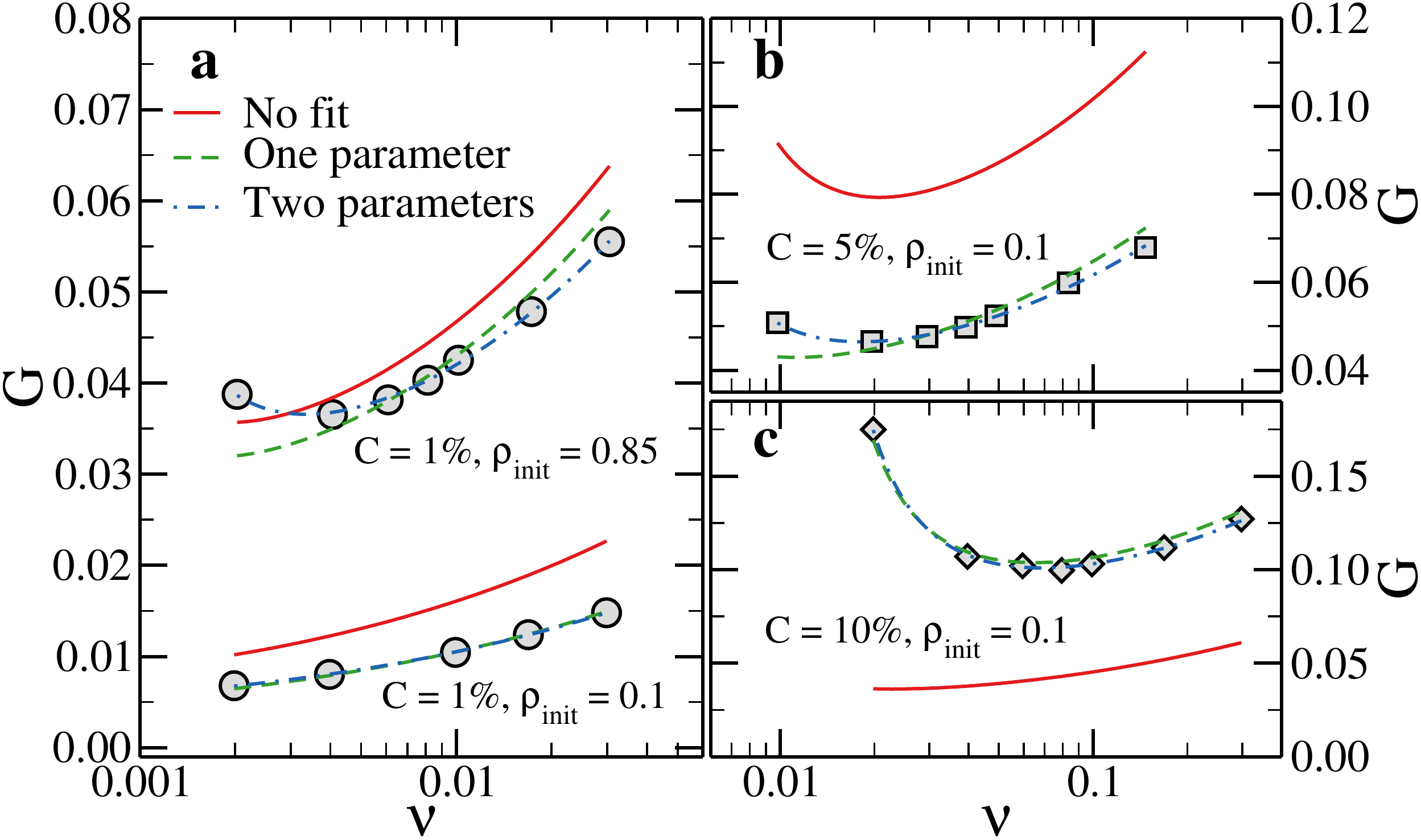}
\includegraphics[width=0.6\textwidth]{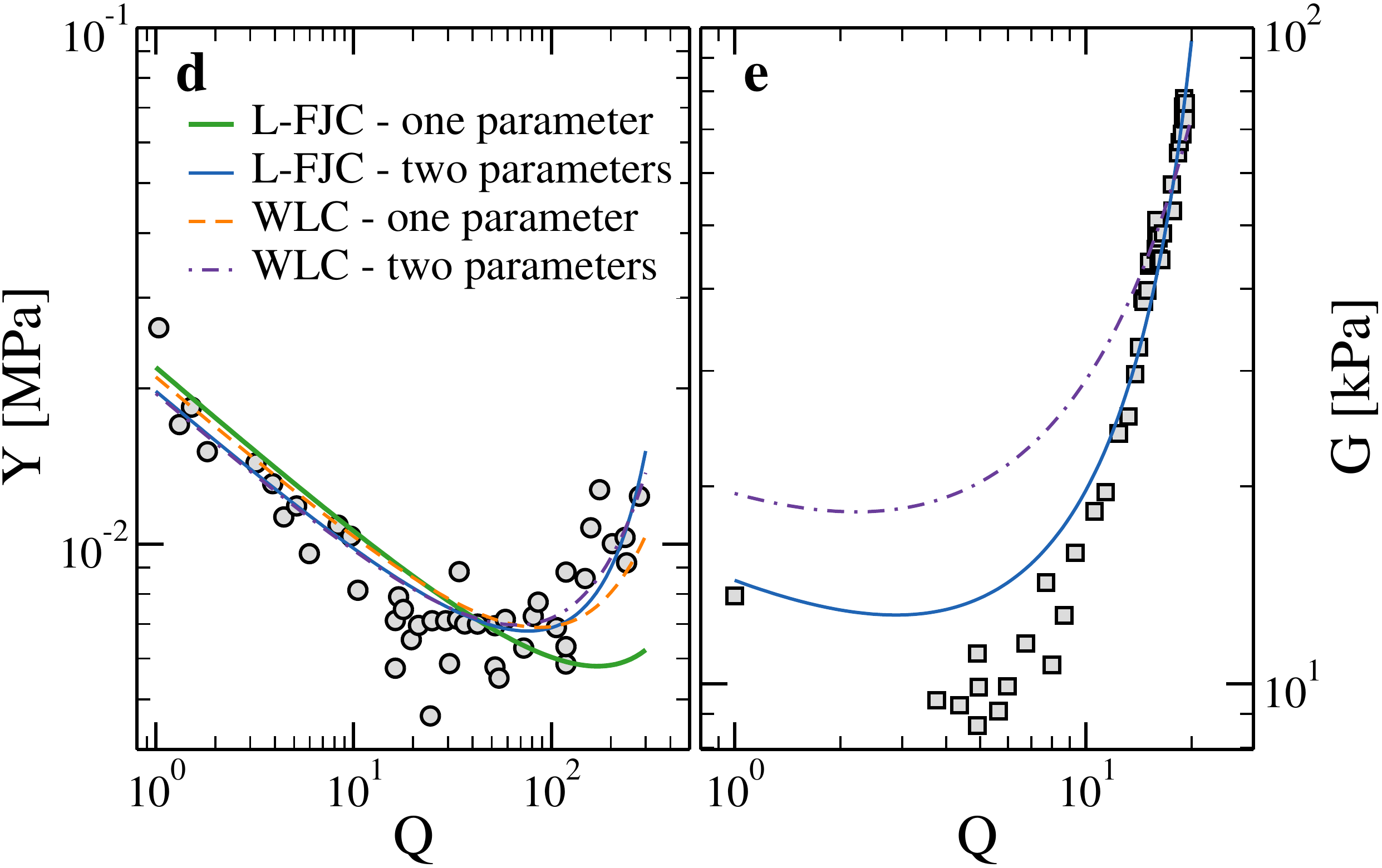}
\caption{\label{fig:fit_results}Fitting results to (\textbf{a}-\textbf{c}) simulation data (\textbf{a}: $C = 1\%$, \textbf{b}: $C = 5\%$, \textbf{c}: $C = 10\%$) and (\textbf{d}-\textbf{e}) experimental data (\textbf{d}: Young modulus taken from \citet{hoshino2018network}, \textbf{e}: shear modulus taken from \citet{matsuda2019fabrication}). The quality of the fits in panel \textbf{e} does not depend on whether the point at $Q=1$ is considered or not or if we restrict the fit to swelling ratios $Q \lesssim 15$.}
\end{figure}

In addition to providing the best comparison with the numerical data in the whole density range, the FJC description also captures the presence and (although only in a semi-quantitative fashion) the position of the minimum. This holds true for all the investigated systems, highlighting the role played by the short chains, whose strong non-Gaussian character heavily influences the overall elasticity of the network.

Although real short chains do not follow the exact end-to-end probability distribution we use here (see Eq.~\eqref{eq:W_FJC}), they are surely far from the scaling regime and hence they should never be approximated as Gaussian chains, even in the melt or close to the theta point. This aspect has important consequences for the analysis of experimental randomly-crosslinked polymer networks, for which one may attempt to extract some microscopic parameter (such as the contour length or the average end-to-end distance) by fitting the measured elastic properties to some theoretical relation such as the ones we discuss here. Unfortunately, such an approach will most likely yield unreliable estimates. This is shown in Figure \ref{fig:fit_results}(a-c), where we compare the $G$ values with the L-FJC model (see Appendix~\ref{sec:models}). Here we use the L-FJC model since we assume, as often done when dealing with similar systems~\cite{hoshino2018network}, that the network can be considered as composed by $N_s$ strands of $\langle n \rangle$ segments. The expression we employ contains two quantities that can be either fixed or fitted to the data: the average end-to-end distance in a specific state (\textit{e.g.} the preparation state), $R_0$, and the average strand length $\langle n \rangle$ (or, equivalently, the contour length $r_{\rm max} = \langle n \rangle b$). Together with the numerical data, in Fig.~\ref{fig:fit_results} we present three sets of theoretical curves: $G$ as estimated by using the values of $R_0$ and $\langle n \rangle$ as obtained from the simulations or fitted by using either $R_0$ or both quantities as free parameters. If $C$ is small (and hence $\langle n \rangle$ is large), the difference between the parameter-free expression and the numerical data is small ($10$ -$15\%$).
However, as $C$ becomes comparable with the values that are often used in real randomly-crosslinked hydrogels ($\approx 5\%$), the difference between the theoretical and simulation data becomes very significant: For instance, for $C = 10\%$ the parameter-free expression fails to capture even the presence of the minimum. Fitting the numerical data makes it possible to achieve an excellent agreement, although the values of the parameters come out to be sensibly different (sometimes more than $50\%$) from the real values (see Sec~\ref{sec:fitting}). 
Our results thus show that even in the simplest randomly-crosslinked system --a phantom network of freely-jointed chains-- neglecting the shortness of the majority of the chains, which dominate the elastic response, can lead to a dramatic loss of accuracy. Randomly-crosslinked polymer networks contain short chains which are inevitably quite far from the scaling regime, and hence even their qualitative behaviour can become elusive if looked through the lens of polymer theories that rest too heavily on the Gaussianity of the chains.

We also apply our theoretical expressions to two sets of data which have been recently published and that exhibit a non-monotonic behaviour. Both experiments have been carried out in the laboratory  of J.~P.~Gong~\cite{hoshino2018network,matsuda2019fabrication}. The first system is a tetra-PEG hydrogel composed of monodisperse long chains that can be greatly swollen by using a combined approach of adding a molecular stent and applying a PEG dehydration method~\cite{hoshino2018network}. Since the system is monodisperse and the chains quite long, we expect the theoretical expressions derived here to work well. Indeed, as shown in Fig.~\ref{fig:fit_results}\textbf{d}, the resulting Young modulus is a non-monotonic function of the swelling ratio $Q$, \textit{i.e.} of the ratio between the volume at which the measurements are performed and the volume at which the sample was synthesised. The experimental data can be fitted with both the L-FJC and worm-like-chain (WLC) expressions (see Sec~\ref{sec:models}), since both models reproduce the data with high accuracy when fitted with two free parameters ($R_0$ and $\langle n \rangle$). However, better results are obtained with the WLC model, which fits well when $\langle n \rangle$ is fixed to its experimentally-estimated value, yielding a value $R_0 = 7.2$ nm, which is very close to the independently-estimated value of $8.1$ nm~\cite{hoshino2018network}, in agreement with what reported in \citet{hoshino2018network}. By contrast, as shown in Fig.~\ref{fig:fit_results}\textbf{e}, the theoretical expressions reported here cannot go beyond a qualitative agreement with experiments of randomly-crosslinked PNaAMPS networks~\cite{matsuda2019fabrication}, even if two parameters are left free and we only fit to the experimental data in a narrow range of swelling ratios. In addition, the fitted values of the two parameters are always unphysical (\textit{e.g.} smaller than one nanometer, see Sec~\ref{sec:fitting}). Although part of the discrepancy might be due to the charged nature of the polymers involved~\cite{fisher1977chain}, we believe that the disagreement between the theoretical and experimental behaviours can be at least partially ascribed to the randomly-crosslinked nature of the network, and hence to the abundance of short chains. Since the end-to-end distribution of such short chains is not known and depends on the chemical and physical details, there is no way of taking into account their contribution to the overall elasticity in a realistic way. These results thus highlight the difficulty of deriving a theoretical expression to assess the elastic behaviour of randomly-crosslinked real networks.

\section{Summary and conclusions}

We have used numerical simulations of disordered phantom polymer networks to understand the role of the chain size distribution on their elastic properties. In order to do so we employed an \textit{in silico} synthesis technique by means of which we can independently control the number and chemical size of the chains, set by the crosslinker concentration, as well as the distribution of their end-to-end distances, which can be controlled by varying the initial monomer concentration. We found that networks composed by chains of equal contour length can have shear moduli that depend strongly on the end-to-end distance even when probed at the same strand concentration. Hence this shows taht even in simple systems the synthesis protocol can have a large impact on the final material properties of the network even when it does not affect the chemical properties of its basic constituents, as recently highlighted in a microgel system~\cite{freeman2020effect}.

We then compared the results from the simulations of the phantom network polymer theory, which was revisited to obtain explicit expressions for the shear modulus assuming three different chain conformation fluctuations, namely the exact freely-jointed chain, Gaussian, and extensible freely-jointed chain models. We observed a non-monotonic behaviour of $G$ as a function of the strand density that, thanks to a comparison with the theoretical results, can be completely ascribed to entropic effects that cannot be accounted for within a Gaussian description. We thus conclude that the role played by short stretched chains in the mechanical description of polymer networks is fundamental and should not be overlooked.
This insight is supported by an analysis of experimental data of the elastic moduli of hydrogels reported in the literature.
We are confident that the numerical and analytical tools employed here can be used to address similar and other open questions concerning both the dynamics and the topology in systems in which excluded-volume effects are also taken into account, and hence entanglements effects may be relevant. Investigations in this direction are under way.

\section*{Appendix} 

\setcounter{section}{0}
\renewcommand{\thesection}{A\Roman{section}}

\section{Shear modulus of a system with Gaussian distributed end-to-end distances}
\label{sec:shear_gauss}

In this section we show how Eq.~\eqref{eq:G_textbook} can be derived for a polydisperse network. Similar derivations for the case of monodisperse networks can be found in standard textbooks \cite{rubinstein2003polymer,mark2007physical}. We start by noting something that is sometimes overlooked: the Gaussian distribution $W_n^G(\mathbf r)$, Eq.~\eqref{eq:W_G}, applies to a single chain.
However, at the ensemble level the distribution of end-to-end vectors, which we may call $\Omega [\mathbf r(n)]$, is not a Gaussian in general. However, if we assume, for example, that the system has been obtained through end-crosslinking starting from a melt of precursor chains~\cite{duering1994structure}, then $\Omega[\mathbf r(n)] = W_n^G(\mathbf r)$ \cite{flory1976statistical}, so that the magnitudes $r$ of the $\mathbf r$ vectors will be Gaussian distributed. Under this assumption, one has 

\begin{equation}
\left \langle \frac{\overline{r^2}}{n b^2} \right \rangle = \left \langle \frac{R^2}{n b^2} \right \rangle + \left \langle \frac{\overline{u^2}}{n b^2} \right \rangle = 1.
\end{equation}

\noindent
To evaluate the term in brackets in Eq.~\eqref{eq:G_gaussian} we thus only need to evaluate the fluctuation term $ \left \langle \frac{\overline{u^2}}{n b^2} \right \rangle $. This term can be computed using the equipartition theorem (the same derivation can be found for example in Ref.~\cite{everaers1998constrained}). The total energy of the fluctuations is $\mathcal U_{\rm fluct} =  \frac 3 2 k_B T N_x$, where $N_x$ is the number of active crosslinkers, since there is one mode for each node and to each mode it is associated an energy of $\frac 3 2 k_BT$ \cite{everaers1998constrained}. Moreover $N_x = 2 N_s/\phi$, with $N_s$ the number of elastically-active strands. Therefore the mean energy per strand is

\begin{equation}
\frac{ \mathcal U_{\rm fluct}}{N_s} = \frac{3 k_B T} \phi = \frac{3 k_B T}{2N_s} \sum_i^{N_s} \frac{\overline{u_i^2}}{n_i b^2} =  \frac {3 k_B T} 2 \left \langle \frac{\overline{u^2}}{n b^2} \right \rangle,
\end{equation}

\noindent
where the sum extends over all the elastically-active strands. Therefore, we get $\left \langle \frac{\overline{u^2}}{n b^2} \right \rangle =  \frac 2 \phi$ (a generalization to the polydisperse case of a well-known result for the phantom network~\cite{graessley1975statistical,higgs1988polydisperse}), from which we finally obtain

\begin{equation}
 \left\langle \frac{R^2}{nb^2} \right\rangle  
 = 1 - \frac 2 \phi.
 \label{eq:textbookondition}
 \end{equation}

\noindent
From Eq.~\eqref{eq:G_gaussian} we obtain Eq.~\eqref{eq:G_textbook}, \textit{i.e.}, $G^G = \left(1-\frac 2 \phi \right) k_B T \nu$.

The validity of Eq.~\eqref{eq:textbookondition} depends not only on the crosslinking procedure but also on the macroscopic thermodynamic parameters (such as solvent quality, density or pressure). For example, if the chain-size distribution is such that short chains are abundant, as it is the case for randomly crosslinked networks \cite{grest1990statistical,higgs1988polydisperse}, the front factor $A$ (see Eq.~\eqref{eq:G_gaussian}) will depend on the chain size distribution, since short chains are non-Gaussian. Another example is when the crosslinking procedure is performed in a state in which the chains are non-Gaussian, \textit{e.g.} under good solvent conditions, where the chains behave as self-avoiding random walks \cite{rubinstein2003polymer}.

\section{Models of chain statistics}
\label{sec:models}

The freely-jointed chain approximation, Eq.~\eqref{eq:W_FJC}, describes an inextensible chain, since each component $f_\alpha, \ \alpha = x,y,z$ of the force $\mathbf f$ required to stretch the chain, \textit{e.g.}

\begin{equation}
f_x = -T \frac{dS_n(\mathbf r)}{dx} = -\frac{k_B T}{W_n(\mathbf r)}\frac{d W_n(\mathbf r)}{dx},
\end{equation}

\noindent
diverges in the limit $r \to nb$, i.e., when the end-to-end distance approaches the contour length. In the limit of large strains and large degree of polymerization $n$, a better approximation is provided by the well-known Langevin dependence of the elongation on the exerted force $f$, which yields for the end-to-end probability distribution function~\cite{treloar1975physics}

\begin{equation}
\label{eq:W_langevin}
W_n^L(\mathbf r) = A \exp\left[-\frac{r}{b} \mathcal{L}^{-1}(r /nb) \right] \left[ \frac{\mathcal{L}^{-1}(r /nb)}{\sinh \mathcal{L}^{-1}(r /nb)} \right]^{-n},
\end{equation}

\noindent
where $\beta = 1 / k_B T$, $T$ is the temperature, $k_B$ is the Boltzmann constant, $A$ is a normalisation constant, and $\mathcal{L}(x)^{-1}$ is the inverse Langevin function. The latter is defined as $\mathcal{L}(x) = \coth{(x)} - 1/x$ and it turns out to be equal to the ratio between the end-to-end distance and the contour length of a chain that is subject to a force $f$:

\begin{equation}
\frac r {nb} =  \coth\left( \frac{fb}{k_BT} \right) - \frac{k_B T }{fb} = \mathcal{L} ( \beta fb ).
\end{equation}

We note on passing that $\mathcal{L}^{-1}(\cdot)$ cannot be written in a closed form~\cite{jedynak2015approximation} and hence must be evaluated numerically.

Phantom Kremer-Grest chains behave exactly as freely-jointed chains up to end-to-end distances that are very close to the contour length. However, beyond those values the Langevin description is no longer valid, and one has to resort to the ex-FJC model, for which the following analytical form of the force-extension curve has been recently derived~\cite{fiasconaro2019analytical}:

\begin{equation}
\label{eq:ex-FJC}
r = nb \mathcal{L} ( \beta fb ) + \frac{nf}{k} \left[ 1 + \frac{1 - \mathcal{L} ( \beta fb ) \coth(\beta f b)}{1 + \frac{f}{k b}\coth(\beta fb)} \right],
\end{equation}

\noindent
where $k$ is the monomer-monomer force constant in the harmonic approximation. In principle it is possible to integrate the inverse of this relation to get the end-to-end probability density. However, as it is clear from Eq.~\eqref{eq:g_ph}, we only need the derivatives of the inverse function, and hence there is no need to obtain $W(r)$ explicitly. We set the value of $k$ to the value of the second derivative of the Kremer-Grest potential as computed in the minimum, $k \approx 867 \, \epsilon / \sigma^2$. We have also fitted the force-extension curves as obtained in simulations of single chains, obtaining values that are compatible with this latter estimate.

We also report the following expression, which approximates the force-elongation relation for a worm-like chain \cite{petrosyan2017improved} (WLC) and is used in the main text to fit the experimental data shown in Fig.~\ref{fig:fit_results}\textbf{c}: 

\begin{equation}
\label{eq:wlc}
\frac{fb}{k_BT} = \frac 1 4 \left(1 -\frac r {nb}\right)^{-2} - \frac 1 4 + \frac r {nb} - 0.8 \left(\frac r {nb} \right)^{2.15}
\end{equation}

\begin{figure}
\includegraphics[width=0.45\textwidth]{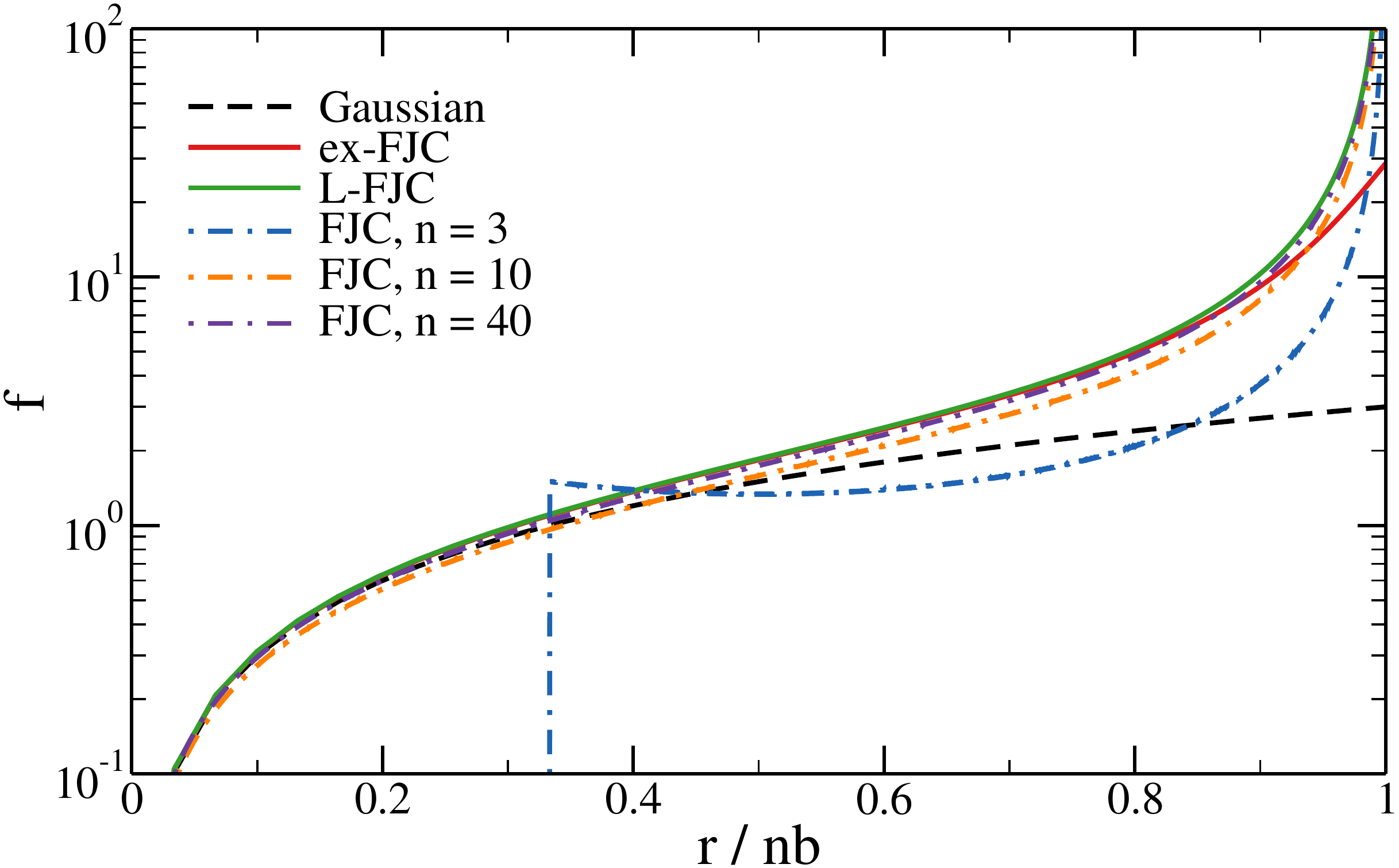}
\caption{\label{fig:force_extensionomparison}The force required to extend a chain by $r$ (scaled by its contour length $nb$) for the Gaussian (dashed line, Eq.~\eqref{eq:W_G}), L-FJC and ex-JFC (solid lines, Eq.~\eqref{eq:ex-FJC}) and FJC (dashed-dotted lines, Eq.~\eqref{eq:W_FJC}) models. Note that plotted in this way $f$ depends on $n$ only for the latter case.}
\end{figure}

Fig.~\ref{fig:force_extensionomparison} shows the force-extension curve for polymer chains described with different models. Since the force is plotted as a function of the distance scaled by the FJC contour length $nb$, the Gaussian, L-FJC and ex-FJC description become independent of $n$. By contrast, the extensional force of the exact FJC model, whose end-to-end probability distribution is given by Eq.~\eqref{eq:W_FJC}, retains an $n$-dependence that is very strong for small $n$ and decreases upon increasing the chain size. Indeed, for $n \gtrsim 40$ the resulting force is essentially $n$-independent and overlaps almost completely with the FJC curve.

\section{Fitting procedure and additional results}
\label{sec:fitting}

As discussed in the main text in Sec.~\ref{sec:results} we fit the experimental Young's and shear moduli reported in \citet{hoshino2018network} (system A) and \citet{matsuda2019fabrication} (system B). As commonly done in the analysis of experimental systems, in both cases we will consider the networks to be formed by strands of average size $\langle n \rangle$ and use relations based on Eq.~\eqref{eq:g_ph}: for system A we use

\begin{equation}
G_{\rm exp} = -\frac{T R^2 \nu}{3} \left[ \frac{ds_{n}(\tilde r)}{d\tilde r} \left( \frac{1}{\tilde r} - \frac{R^2}{2  \tilde r^3} \right) + \frac{d^2 s_{n}(\tilde r)}{d \tilde r^2} \frac{R^2}{2  \tilde r^2} \right]
\end{equation}

\noindent while for system B we fit  $Y_{\rm exp}=3 G_{\rm exp}$. We fit the experimental data with the L-FJC and WLC models by using Eqs.~\eqref{eq:W_langevin} and ~\eqref{eq:wlc} to numerically evaluate the derivatives of $s_{n}(\tilde r)$.

We evaluate $R$, $\tilde{r}$ and $\nu$ as follows. Let $V_0$, $\nu_0$ and $R_0$ be the volume, chain number density and average end-to-end distance of the polymer chains in the preparation state and $V$, $\nu$ and $R$ the same quantities for the generic state point at which the experimental measurements are carried out. We define the swelling ratio $Q = V / V_0 = \nu_0 / \nu$ so that $\nu = \nu_0 / Q$ and $R = \left( \frac{V}{V_0} \right)^{1/3} R_0 = Q^{1/3} R_0$. Since $\tilde{r}^2 = \overline{r^2} = R^2 + \overline{u^2}$ and, as shown in Sec~\ref{sec:shear_gauss}, $\overline{u^2} = \frac{2}{\phi} nb^2$, for systems with $\phi = 4$ we also have that $\tilde{r} = \sqrt{R^2 + r_{\rm max} b / 2}$, 
where $r_{\rm max} = \langle n \rangle b$ is the contour length of the strands. Using the numbers reported in the original papers, we find that $\nu_0^A = 1.97 \cdot 10^{-3}$ nm$^{-3}$ and $\nu_0^B = 0.245$ nm$^{-3}$. For system A the authors also report independent estimates for $R_0$ ($R_0^A = 8.1$ nm) and $r_{\rm max}$ ($r_{\rm max}^A = 82$ nm). For system A we show fitting results obtained by either fixing $r_{\rm max}$ and using $R_0$ as a fitting parameter or by leaving both quantities as fitting parameters. For system B we do the latter.

For system A we obtain $R_0^{\rm FJC} = 7.46$ nm and $R_0^{\rm WLC} = 7.2$ nm for the one-parameter fits and $R_0^{\rm FJC} = 5.08$ nm, $r_{\rm max}^{\rm FJC} = 42.75$ nm and $R_0^{\rm WLC} = 7.2$, $r_{\rm max}^{\rm WLC} = 63.87$ nm for the two-parameter fits.

For system $B$ we obtain unphysical values ($R_0^{\rm FJC} = 0.11$ nm, $r_{\rm max}^{\rm FJC} = 0.65$ nm and $R_0^{\rm WLC} = 0.17$, $r_{\rm max}^{\rm WLC} = 0.96$ nm). The results do not improve if we restrict the fitting range to narrower $Q$-ranges.

\begin{acknowledgement}

We thank F. Sciortino and D. Truzzolillo for helpful discussions and J. P. Gong, T. Nakajima and T. Matsuda for sharing the data reported in Fig.~\ref{fig:fit_results}d-e.
We acknowledge financial support from the European Research Council (ERC Consolidator Grant 681597, MIMIC). W. Kob is senior member of the Institut universitaire de France.
\end{acknowledgement}


\bibliography{bibliography.bib}

\section{Supplemental Material}

\subsection{Shear modulus in the affine network model}
\label{sec:affine}

Eq.~8 was obtained under the phantom network assumption, \textit{i.e.}, that the coordinates of the vector $\mathbf r_\lambda = \mathbf R_\lambda + \mathbf u_\lambda$ transform according to $R_{x,{\lambda} }= \lambda R_{x,1}$ and $u_{x,\lambda} = u_{x,1}$. In this case, the fluctuation term is unaffected by the deformation. One can also assume, on the contrary, that the fluctuations deform affinely with the average end-to-end vector, \textit{i.e.},  $r_{x,{\lambda}} = \lambda r_{x,1}$ and analogous for the other coordinates. In this case, one obtaines

\begin{equation}
\label{eq:g_aff}
g^{\rm aff} = -\frac{T R^2}{6 V} \left[ \frac{ds_{n}(\tilde r)}{d \tilde r} \frac{1}{\tilde r} +  \frac{d^2 s_{n}(\tilde r)}{d \tilde r^2} \right],
\end{equation}

\noindent
which results in a different Gaussian modulus:

\begin{equation}
\label{eq:G_gaussian_affine}
G^{\rm{aff}, G} = \frac{k_B T}{V} \sum_1^{N_s} \frac{\overline{r_i^2}}{n_ib^2} = \left\langle \frac{\overline{r^2}}{nb^2} \right\rangle k_B T \nu \equiv A^{\rm aff} k_B T \nu,
\end{equation}

\noindent
where the sum is, as usual, taken over the $N_s$ elastically-active strands. Under the assumption that the $\overline{r_i^2}$ are Gaussianly distributed (see Sec.~AI in the main text) we have

\begin{equation}
\left \langle \frac{\overline{r^2}}{n b^2} \right \rangle = 1.
\label{eq:textbookondition_aff}
\end{equation}

\noindent
From Eqs.~\eqref{eq:textbookondition_aff} and ~\eqref{eq:G_gaussian_affine}, get the commonly reported expression~\cite{rubinstein2003polymer,mark2007physical}

\begin{equation}
\label{eq:G_textbook_aff}
G^{\rm{aff},G} = k_B T \nu.
\end{equation}

\subsection{Monomer mean-squared displacement during equilibration}
\label{sec:msd}

\begin{figure}
\includegraphics[width=0.5\textwidth]{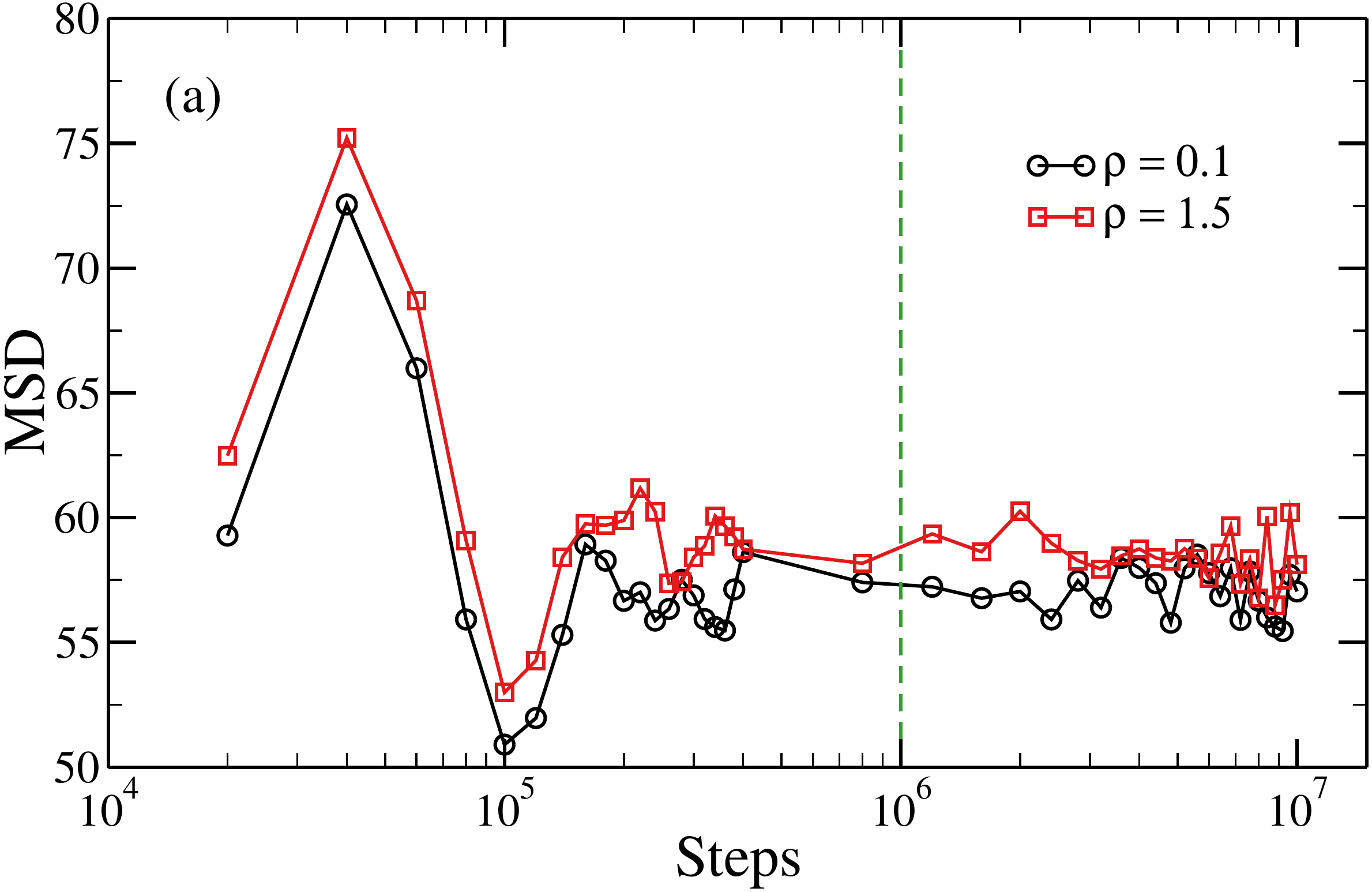}
\caption{\label{fig:msd} The mean-squared displacement of the $C = 1\%$, $\rho_{\rm init} = 0.1$ sample computed at $\rho = 0.1$ and $\rho = 1.5$. The vertical dashed line signals the equilibration time we use.}
\end{figure}

In order to verify that the system has equilibrated correctly, we measure the monomer mean-squared displacement (MSD). In Fig.~\ref{fig:msd} we report the monomer MSD of the $C = 1\%$, $\rho_{\rm init} = 0.1$ sample computed at $\rho = 0.1$ and $\rho = 1.5$. We note that the MSD quickly reaches a plateau, signaling that the oscillation modes of all the strands have equilibrated.

\subsection{Density scaling of RMS equilibrium end-to-end distance}
\label{sec:rho_scaling}

\begin{figure}
\includegraphics[width=0.43\textwidth]{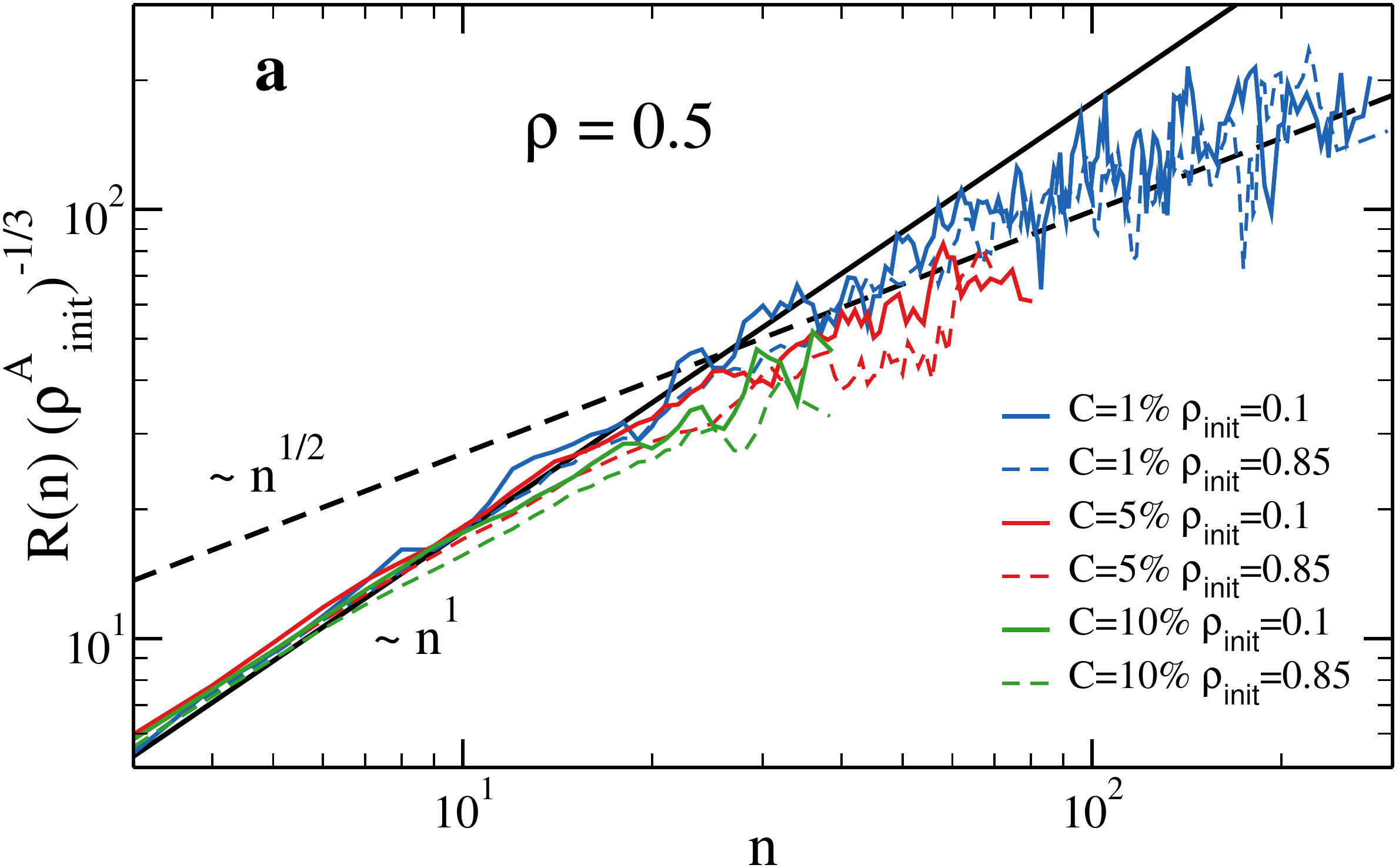}
\includegraphics[width=0.43\textwidth]{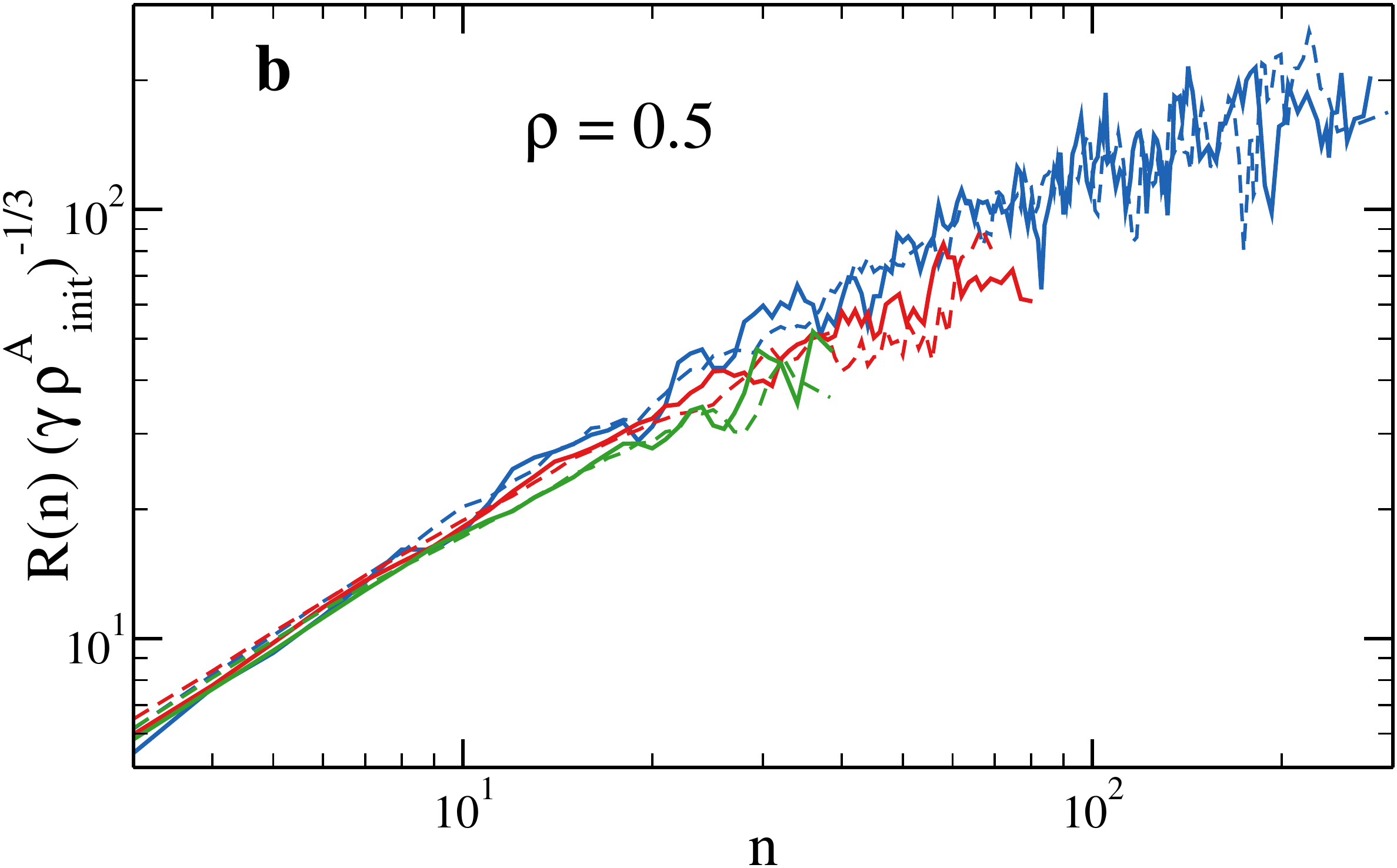}
\includegraphics[width=0.43\textwidth]{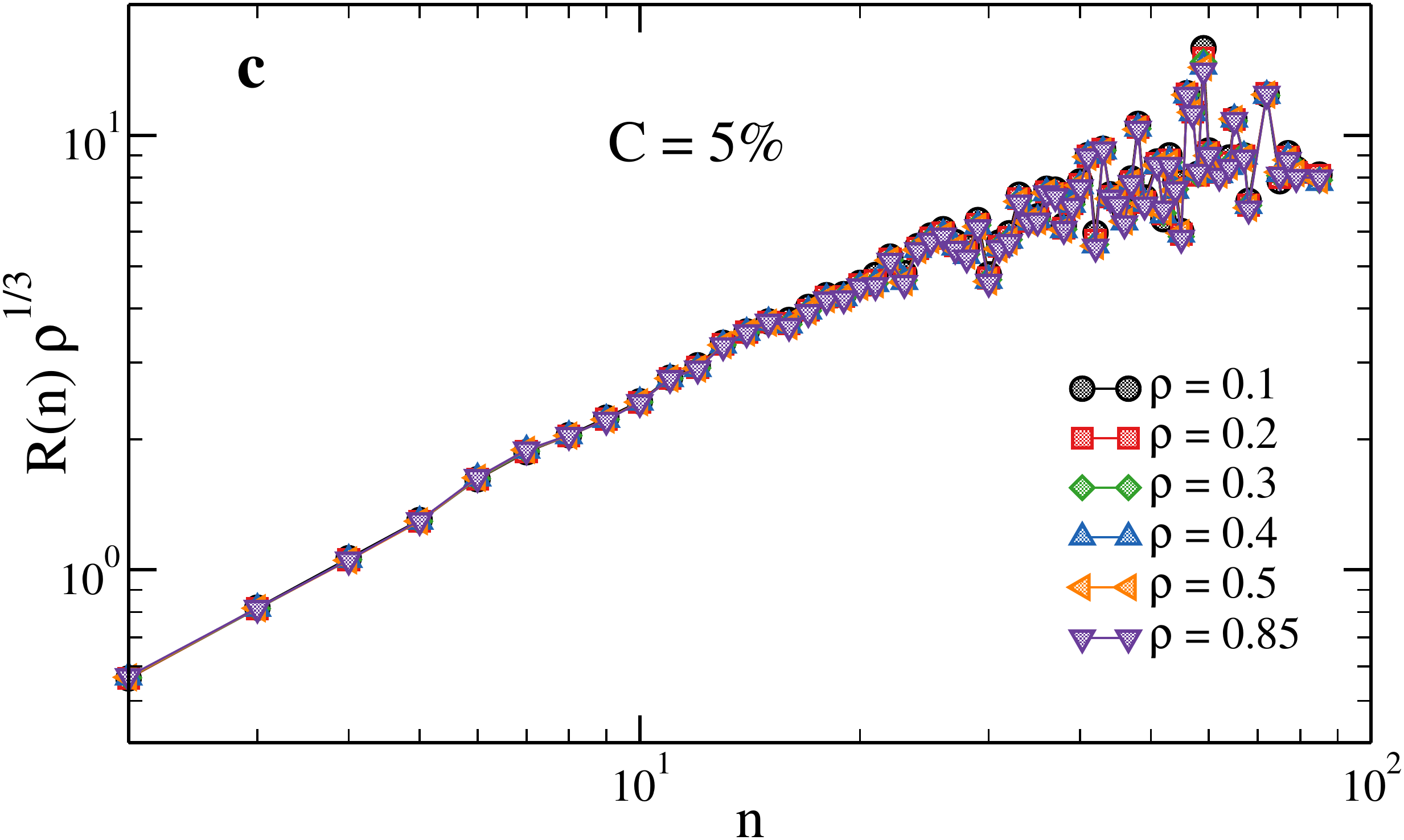}
\caption{\label{fig:RofN} (\textbf{a}-\textbf{b}): RMS equilibrium end-to-end distance of the strands for $C=1\%,5\%,10\%$ for $\rho_{\rm init} =0.1,85$ and $\rho =0.5$, rescaled by (\textbf{a}) the inverse of the initial average distance between neighboring crosslinkers $(\rho^{A}_{\rm init})^{1/3}=(C \rho_{\rm init})^{1/3}$ and (\textbf{b}) by $(\gamma \rho^{A}_{\rm init})^{1/3}$, where $\gamma = 0.74$ for $\rho_{\rm init} = 0.85$ and $\gamma = 1$ for $\rho_{\rm init} = 0.1$. (\textbf{c}): Same quantity as in \textbf{a}-\textbf{b}, rescaled by  $\rho^{1/3}$, for $C=5\%$ and for different values of $\rho$.}
\end{figure}

We report in Fig.~\ref{fig:RofN}\textbf{a} for $\rho=0.5$ the RMS equilibrium end-to-end distance of the strands, defined as $R(n) \equiv [\langle R^2(n)\rangle_n]^{1/2}$ \footnote{We recall that the end-to-end distance is $\mathbf r(t) \equiv R + \mathbf u(t)$, and that $\overline{r^2} = R^2 + \overline{u^2}$.}, where $\langle \cdot \rangle_n$ denotes the average over all the strands of length $n$. We note that curves for different initial densities $\rho_{\rm init}$ and crosslinker concentrations $C$ fall on the same master curve if divided by the quantity $(\rho^{\rm cl}_{\rm init})^{1/3}$, where $\rho^{\rm cl}_{\rm init}= C \rho_{\rm init}$ is the initial crosslinker density. Since this quantity represents the inverse of the initial average distance between neighboring crosslinkers, we can conclude that the initial spatial distribution of the crosslinkers completely controls the equilibrium end-to-end distance of the chains in the final state. An even better collapse can be obtained by using slightly different (heuristic) factors for the two values of the initial density we use here: Fig.~\ref{fig:RofN}\textbf{b} shows the same curves rescaled by $(\gamma \rho^{A}_{\rm init})^{1/3}$, where $\gamma = 0.74$ for $\rho_{\rm init} = 0.85$ and $\gamma = 1$ for $\rho_{\rm init} = 0.1$.
We note that the same rescaling does not apply to $[\langle \overline{r^2(n)}\rangle_n]^{1/2}$, since the fluctuation term $\overline{u^2}$ does not follow this scaling. We also report in Fig.~\ref{fig:RofN}\textbf{a} the scaling behavior expected for Gaussian strands, i.e., $R(n) \propto n^{1/2}$ (dashed line), and the one for stretched strands, i.e., $R (n) \propto n$ (solid line). One can see that the short chains are on average stretched, and only for larger values of $n$ the Gaussian behavior is recovered. Finally, we remark that since the equilibrium end-to-end distances deform affinely with the network, $R(n)$ curves at different final densities $\rho$ collapse on the same master curve when multiplied by $\rho^{1/3}$, as shown in Fig.~\ref{fig:RofN}\textbf{c}.

\end{document}